\documentclass[fleqn,usenatbib]{mnras}
\usepackage{newtxtext,newtxmath}
\usepackage[T1]{fontenc}
\DeclareRobustCommand{\VAN}[3]{#2}
\let\VANthebibliography\thebibliography
\def\thebibliography{\DeclareRobustCommand{\VAN}[3]{##3}\VANthebibliography}
\usepackage{graphicx}
\usepackage{amsmath}
\usepackage{tabularx}
\usepackage{enumitem,booktabs}
\usepackage[referable]{threeparttablex}
\renewlist{tablenotes}{enumerate}{1}
\makeatletter
\setlist[tablenotes]{label=\tnote{\alph*},ref=\alph*,itemsep=\z@,topsep=\z@skip,partopsep=\z@skip,parsep=\z@,itemindent=\z@,labelindent=\tabcolsep,labelsep=.2em,leftmargin=*,align=left,before={\footnotesize}}
\makeatother
\title[New insight into the massive eccentric binary HD\,165052]{New insight into the massive eccentric binary HD\,165052\thanks{Based on optical spectra collected at the European Southern Observatory with the TIGRE telescope (La Luz, Mexico).}: self-consistent orbital solution, apsidal motion, and fundamental parameters}
\author[S. Rosu et al.]{S.\ Rosu$^{1,2}$\thanks{srosu@kth.se}, E.\ A.\ Quintero$^{3,4}$, G.\ Rauw$^{2}$, and P.\ Eenens$^{3}$
\\
$^{1}$Department of Physics, KTH Royal Institute of Technology, The Oskar Klein Centre, AlbaNova, SE-106 91 Stockholm, Sweden\\
$^{2}$Space sciences, Technologies and Astrophysics Research (STAR) Institute, Universit\'e de Li\`ege, All\'ee du 6 Ao\^ut, 19c, B\^at B5c, 4000 Li\`ege, Belgium\\
$^{3}$Departamento de Astronom\'ia, Universidad de Guanajuato, Guanajuato, Mexico\\
$^{4}$Observatorio Astron\'omico (OAUTP), Universidad Tecnol\'ogica de Pereira, Risaralda, Colombia}
\date{Accepted 2023 March 13. Received 2023 March 6; in original form 2023 January 17}
\pubyear{2023}

\begin{document}
\label{firstpage}
\pagerange{\pageref{firstpage}--\pageref{lastpage}}
\maketitle

\begin{abstract}
HD\,165052 is a short-period massive eccentric binary system that undergoes apsidal motion. As the rate of apsidal motion is directly related to the internal structure constants of the binary components, its study allows to get insight into the internal structure of the stars.
We use medium- and high-resolution spectroscopic observations of HD\,165052 to provide constraints on the fundamental properties of the binary system and the evolutionary state of its components.
We apply a spectral disentangling code to reconstruct artefact-free spectra of the individual stars and derive the radial velocities (RVs) at the times of the observations. We perform the first analysis of the disentangled spectra with the non-LTE model atmosphere code \texttt{CMFGEN} to determine the stellar properties. We derive the first self-consistent orbital solution of all existing RV data, including those reported in the literature, accounting for apsidal motion. We build, for the very first time, dedicated stellar evolution tracks with the \texttt{Cl\'es} code requesting the theoretical effective temperatures and luminosities to match those obtained from our spectroscopic analysis. The binary system HD\,165052, consisting of an O6.5\,V((f)) primary ($T_\text{eff,P}=37\,500 \pm 1000$\,K) and an O7\,V((f)) secondary ($T_\text{eff,S}=36\,000 \pm 1000$\,K), displays apsidal motion at a rate of $(11.30^{+0.64}_{-0.49})^\circ$\,yr$^{-1}$. Evolutionary masses are compared to minimum dynamical masses to constrain the orbital inclination. 
Evolutionary masses $M_\text{ev,P}=24.8\pm1.0\,M_\odot$ and $M_\text{ev,S}=20.9\pm1.0\,M_\odot$ and radii $R_\text{ev,P}=7.0^{+0.5}_{-0.4}\,R_\odot$ and $R_\text{ev,S}=6.2^{+0.4}_{-0.3}\,R_\odot$ are derived, and the inclination is constrained to $22.1^\circ \le i \le 23.3^\circ$. Theoretical apsidal motion rates, derived assuming an age of $2.0\pm0.5$\,Myr for the binary, are in agreement with the observational determination. The agreement with theoretical apsidal motion rates enforces the inferred values of the evolutionary stellar masses and radii.
\end{abstract}


\section{Introduction}\label{sect:intro}
\begin{table*}
\centering
\caption{Physical and orbital parameters of HD\,165052 from the literature.}
\begin{tabular}{llllll}
\hline\hline
\vspace{-3mm}\\
Reference& \citet{stickland97} & \citet{arias02} & \citet{linder07} & \citet{FER} \\
Parameter & \multicolumn{4}{c}{Value} \\
\hline
\vspace{-3mm}\\
Spectral type & ... & O6.5\,V + O7.5\,V & O6\,V + O6.5\,V & O7\,Vz + O7.5\,Vz\\
$P_\text{orb}$ (d) &  $2.955055\pm0.000010$  & $2.95510\pm 0.00001$ & $2.95515\pm0.00004$ & $2.95506\pm0.00002$ \\
$e$ &   0 (fixed)& $0.090\pm0.004$  & $0.081\pm0.015$ &  $0.090\pm0.003$ \\
$K_\text{P}$ (km\,s$^{-1}$) & $95.6\pm2.2$ & $94.8\pm0.5$ & $96.4\pm1.6$ & $97.4\pm0.4$  \\
$K_\text{S}$ (km\,s$^{-1}$) & $109.6\pm2.2$ & $104.7\pm0.5$ & $113.5\pm1.9$ & $106.5\pm0.4$ \\
$i$ ($^\circ$) & $18.7\pm0.4$ (approx.) & ... & ... &  ... \\
$a_\text{P}\sin i$ ($R_\odot$) & $5.58\pm0.13$ & $5.51\pm0.03$ & $5.6\pm 0.1$ & $5.66\pm 0.03$ \\
$a_\text{S}\sin i$ ($R_\odot$) & $6.40\pm0.13$ & $6.09\pm0.03$ & $6.6\pm0.1$ & $6.20\pm0.03$\\
$M_\text{P}\sin^3 i$ ($M_\odot$) & $2.23\pm0.06$ & $1.26\pm0.03$ & $1.5\pm0.1$ & $1.34\pm0.03$ \\
$M_\text{S}\sin^3 i$ ($M_\odot$) & $1.41\pm0.07$ & $1.14\pm0.03$ & $1.3\pm0.1$  & $1.22\pm0.03$\\
$q=M_\text{S}/M_\text{P}$ & $0.873\pm0.027$ & $0.90\pm0.01$ & $0.85\pm0.01$ & $0.91\pm0.01$\\
$R_\text{P}$ ($R_\odot$) & $15.5\pm1.5$ & ... & ...&  ...\\
$R_\text{S}$ ($R_\odot$) &  $14.6\pm 1.1$ & ... & ... & ...\\
\vspace{-3mm}\\
\hline
\end{tabular}
\begin{tablenotes}
\item The parameters are the following: $P_\text{orb}$, the orbital period of the system; $e$, the eccentricity of the orbit; $K_\text{P}$ (respectively $K_\text{S}$), the amplitude of the RV curve of the primary (respectively secondary) star; $i$, the orbital inclination; $a_\text{P}\sin i$ (respectively $a_\text{S}\sin i$), the projected semi-major axis of the primary (respectively secondary) orbit; $M_\text{P}\sin^3 i$ (respectively $M_\text{S}\sin^3 i$), the minimum mass of the primary (respectively secondary) star; $q=M_\text{S}/M_\text{P}$, the mass ratio of the system; and $R_\text{P}$ (respectively $R_\text{S}$), the radius of the primary (respectively secondary) star. The errors represent $\pm 1\sigma$.
\end{tablenotes}
\label{table:intro}
\end{table*} 

Massive stars play a key role in many processes in the Universe, notably through their winds and powerful supernova explosions that contribute to the chemical enrichment of the Universe.  It is nowadays thought that the majority of the massive stars (i.e. O and B-type stars having a mass larger than $8\,M_\odot$) are, or have been, part of a binary or higher multiplicity system \citep{Duchene13}. According to current estimates, more than 70\% of the massive stars inhabiting our Galaxy have spent some part of their existence in a binary system \citep{sana12}. The simple fact of being bound to another star by gravitational attraction can deeply modify the evolutionary track of the considered star, as its evolution is now driven not only by its own properties -- initial mass, mass-loss rate (a problem that has not been self-consistently solved in stellar structure and evolution models yet), and rotation rate \citep{ekstrom12, brott11} --, but also by the binary orbit's parameters \citep[e.g.,][]{wellstein01}. The study of  detached eclipsing and/or spectroscopic binaries is the most accurate way to obtain reliable physical properties of O and B stars. Among these systems, those showing evidence for apsidal motion (i.e. slow precession of the binary system major axis) offer additional possibilities to sound the interior of these massive stars. Indeed, the apsidal motion rate is directly related to the internal structure constant of the stars that make up the binary system \citep{shakura}. The internal structure constant is a sensitive indicator of the density stratification inside a star and its value strongly changes as the star evolves away from the main sequence. Measuring the rate of apsidal motion in a binary system hence not only provides a diagnostic of the -- otherwise difficult to constrain -- internal structure of the stars, but also offers a test of our understanding of stellar structure and evolution \citep{claret10, mazeh08}.

The massive eccentric binary HD\,165052 is a well-known early-type double-lined spectroscopic (SB2) binary system of the very young open cluster NGC\,6530. NGC\,6530 is itself located in the centre of the H\,{\sc ii} region M8, also known as the Lagoon Nebula. This cluster is one of the most studied clusters in our Galaxy, notably for its interesting history of ongoing star formation \citep{sung00, arias02}. The age of NGC\,6530 was estimated by several authors \citep[we refer to][for a critical discussion]{FER}. Most authors agree about an age estimate around 2\,Myr: \citet{vanaltena72} and \citet{mayne08} both derived an age of 2\,Myr, \citet{sagar78} set a lower limit of 2\,Myr though star formation would have continued as recently as 0.25\,Myr ago,  \citet{kilambi77} estimated the age to range from 1 to 3\,Myr, \citet{sung00} estimated the age to 1.5\,Myr with an age spread of 4\,Myr, \citet{damiani04} derived a median age of cluster stars in the central region of 0.8\,Myr, with a maximum age spread for the whole NGC\,6530 cluster of 4\,Myr, and \citet{prisinzano05} derived an age of 2.3\,Myr with an age spread compatible with 2\,Myr. Only \citet{boehm84} derived a much larger age of $5\pm2$\,Myr.

\citet{plaskett24} was the first to report evidence for binarity of HD\,165052 through the discovery of a variable radial velocity (RV). \citet{conti74} pointed out the double-lined binary nature of HD\,165052. The very first orbital solution was proposed by \citet{morrison78}. However, due to the apparent similarity in the optical region between the two components of the binary, the authors did not unambiguously identify the two components and derived an orbital period of 6.14\,days that appeared to be erroneous. \citet{stickland97} revisited the orbital period $P_\text{orb}=2.96$\,days of the system, among other properties (see Table\,\ref{table:intro}), based on 15 high-resolution \textit{IUE} spectra. The binary system was classified as O6.5\,V + O6.5\,V by \citet{walborn72}, modified to O6.5\,V(n)((f)) + O6.5\,V(n)((f)) by \citet{walborn73}, and to O6\,V + O6\,V by \citet{penny96}. \citet{arias02} presented a new optical spectroscopic study of HD\,165052 based on intermediate and high-resolution CCD observations. The authors derived a new orbital solution for the system (see Table\,\ref{table:intro}) and found, for the first time, evidence of apsidal motion in the system from the comparison with previous RV determinations \citep{arias02}. \citet{linder07} investigated the Struve-Sahade effect that was originally pointed out in the system by \citet{arias02} and also derived a new orbital solution for the system (see Table\,\ref{table:intro}). Finally, \citet{FER} presented a new set of RV measurements of HD\,165052 which they obtained through the disentangling of high-resolution optical spectra using the method described by \citet{GL}. They provided a new orbital solution for the system (see Table\,\ref{table:intro}) and confirmed the variation of the longitude of periastron with time. The authors derived the first estimate of the apsidal motion rate in the system: $\dot\omega=(12.1 \pm 0.3)^\circ$\,yr$^{-1}$. 

In this article, we perform a new, in-depth spectroscopic investigation of the binary HD\,165052 through the analysis of both old and new medium- and high-resolution spectra. We reassess the fundamental and orbital parameters of the system making use of the most advanced disentangling method proposed by \citet{quintero20} that allows the reconstruction of artefact-free individual spectra. We further analyse those reconstructed spectra with, for the first time, a non-LTE model atmosphere code, namely \texttt{CMFGEN}. We derive the first self-consistent orbital solution of all existing RV data, including those reported in the literature, accounting for the change of the longitude of periastron with time. Given that the rate of apsidal motion is directly related to the internal stellar structure constants of the binary components, its determination allows us to infer additional constraints to perform critical tests of stellar structure and evolution models \citep[and references therein]{claret19, claret21}. Finally, we also compute dedicated stellar structure and evolution tracks to derive evolutionary masses and radii for the stars and to put a robust constraint on the value of the inclination of the orbit, as well as to see how the theoretical apsidal motion rates compare to the observational value. 

The set of spectroscopic data we use is introduced in Sect.\,\ref{sect:observations}. In Sect.\,\ref{sect:specanalysis} we perform the spectral disentangling, reassess the spectral classification of the stars, and analyse the reconstructed spectra by means of the {\tt CMFGEN} non-LTE model atmosphere code \citep{Hillier}. The RVs deduced from the spectral disentangling are combined with data from the literature in Sect.\,\ref{sect:omegadot} to derive values for the orbital period, the mass ratio, the rate of apsidal motion, and the orbital eccentricity of the system, among others. The stellar structure and evolution tracks computed with the Code Li\'egeois d'\'Evolution Stellaire are presented in Sect.\,\ref{sect:cles}. We provide our conclusions in Sect.\,\ref{sect:conclusion}.

\section{Observational data}\label{sect:observations}
\subsection{Spectroscopy}
We extracted a total of 47 medium- and high-resolution \'echelle spectra of HD\,165052 in the optical domain from the ESO science archive. Those spectra were collected between May 1999 and July 2017 using different instruments. Twenty-one spectra were obtained with the Fiber-fed Extended Range Optical Spectrograph (FEROS) mounted on the European Southern Observatory (ESO) 1.5\,m telescope in La Silla, Chile \citep{Kaufer}, between May 1999 and April 2002, three spectra were obtained with the FEROS instrument mounted on the ESO 2.2\,m telescope between May 2004 and April 2007. FEROS has a spectral resolving power of 48\,000. The FEROS data were reduced using the FEROS pipeline of the {\tt MIDAS} software. Ten spectra were obtained with the ESPaDOns spectrograph attached to the Canada-France-Hawa\"i observatory (CFH) 3.6\,m telescope in Hawa\"i \citep{donati03}. These data were collected during a single night in June 2010. ESPaDOnS has a spectral resolving power of 68\,000. The reduced ESPaDOnS data were retrieved from the CFHT archive. Four spectra were obtained with the Ultraviolet and Visual Echelle Spectrograph (UVES) mounted on the ESO Very Large Telescope (VLT) at Cerro Paranal, Chile \citep{dekker00}, during a single night in November 2002. UVES has a spectral resolving power of 65\,030 in the blue arm and of 74\,450 in the red arm. The wavelength domain ranges from 3000 to 5000\,$\AA$ (blue arm) and from 4200 to 11\,000\,$\AA$ (red arm). Nine spectra were obtained with the XSHOOTER instrument mounted on the VLT \citep{vernet11}, in July 2016 and July 2017. The wavelength domain ranges from 3000 to 5595\,$\AA$ (UVB) and from 5595 to 10\,240\,$\AA$ (VIS). The spectral resolving power of XSHOOTER is, depending on the observations, 4112 or 5453 (UVB) and 6505, 8935, or 11\,333 (VIS). The ESPaDOns, UVES, and XSHOOTER spectra provided in the archive were already reduced using the dedicated pipelines. We complemented those spectra with 24 high-resolution \'echelle spectra obtained with the HEROS instrument mounted on the TIGRE telescope between April 2019 and April 2021 \citep{gonzalez22}. The wavelength domain ranges from 3500 to 5600\,$\AA$ (blue channel) and from 5800 to 8800\,$\AA$ (red channel). The spectral resolving power is 20\,000. For all the spectra, we removed cosmic rays and telluric absorption lines using {\tt MIDAS} and the {\tt telluric} tool within {\tt IRAF}, respectively.  We normalised the spectra with {\tt MIDAS} by fitting low-order polynomials to the continuum as in \citet{rosu20b, rosu22a, rosu22b}. The journal of the spectroscopic observations is presented in Table\,\ref{Table:spectro+RV}, together with the radial velocities computed in Sect.\,\ref{subsect:spectraldisentangling} and the phase computed using the orbital period derived in Sect.\,\ref{sect:omegadot}.

\section{Spectral analysis}\label{sect:specanalysis}
\subsection{Spectral disentangling\label{subsect:spectraldisentangling}}
As a first step, we performed the spectral disentangling of all data using our disentangling code based on the method described by \citet{GL}. We derived the individual spectra of the binary components as well as their RVs at the times of the observations. We refer to \citet{GL}, \citet{marchenko98}, and \citet{quintero20} for detailed information about the methodology and its limitations, and to Appendix\,\ref{appendix:spectrotable} for a detailed description of the method adopted in the present context.

The resulting RVs of both stars are reported in Table\,\ref{Table:spectro+RV} together with their $1\sigma$ errors.

We confirm the observations of \citet{FER} who reported artefacts in the spectra of the stellar components of HD\,165052 after disentangling using the version of the ``shift and add'' technique of \citet{marchenko98}, a precursor of the code proposed by \citet{GL}. \citet{quintero20} demonstrated that the shift and add method produces artefacts when the spectra contain broad lines with low Doppler shifts. These artefacts distort the spectral line profiles, producing uncertainties in the flux and the profile of the spectral lines of the stellar components.

The novel QER20 spectral disentangling algorithm proposed by \citet{quintero20} combines the advantages of the shift and add method (versatility and ease of implementation) with artefact-free reconstructed spectra. The fundamental principle of the QER20 Package consists in considering the integrated flux of a given spectral line as a free parameter \citep[see][for a detailed description of the QER20 algorithm]{quintero20}.

\begin{figure}
\centering
\includegraphics[clip=true, trim=0 5 20 15,width=\linewidth]{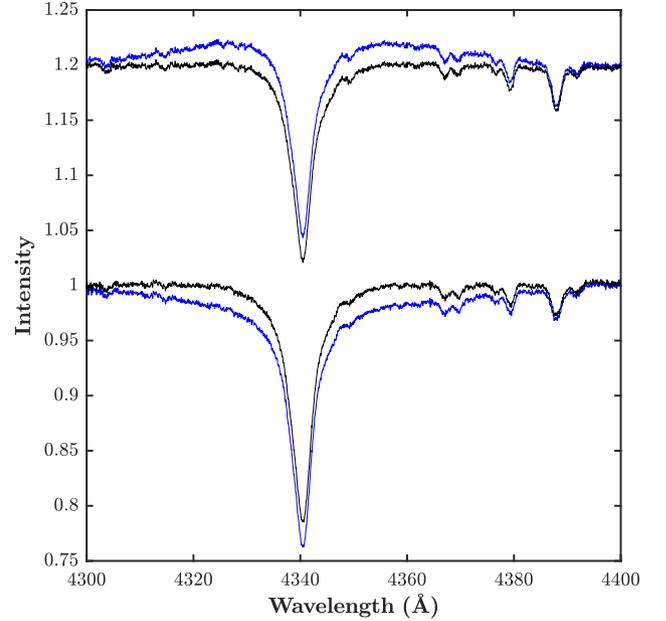}
\caption{Reconstructed spectra of the stellar components of HD\,165052 in the wavelength domain around the H$\gamma$ line, using the shift and add method (blue) and the QER20 Package (black). The spectra of the stellar components are vertically shifted for clarity.\label{fig:QER}}
\end{figure}

Hence, as a second step, we used the QER20 algorithm, fixing the RVs of the stars to those computed previously and reported in Table\,\ref{Table:spectro+RV}. In order to illustrate the reliability of this novel disentangling method in the case of the binary HD\,165052, Figure\,\ref{fig:QER} compares the reconstructed H$\gamma$ line of both stellar components, obtained with the shift and add method and the QER20 Package. The former shows the artefacts mentioned above: emission wings in the spectrum of the primary, and absorption wings in the secondary. In contrast, the QER20 Package yields profiles free of these artefacts. The difference in integrated flux of the lines amounts to 5\% to 7\% between the two methods.

\subsection{Spectral classification and absolute magnitudes}\label{subsect:spectralclass}
We reassessed the spectral classification of the binary components of HD\,165052 based on their reconstructed spectra. 

To determine the spectral types of the stars, we used Conti's criterion \citep{Conti71} complemented by \citet{Mathys88}. We found that $\log W' = \log[\text{EW(He\,{\sc i} $\lambda$\,4471)/EW(He\,{\sc ii} $\lambda$\,4542)}]$ amounts to $-0.17\pm 0.01$ for the primary star and to $-0.05 \pm 0.01$ for the secondary star, which correspond to spectral types O6.5 and O7 (with spectral type O7.5 within the error bars), respectively. We also followed the Walborn criterion \citep{walborn90} based on the strengths of the He\,{\sc ii} $\lambda$\,4686 line and the N\,{\sc iii} $\lambda\lambda$\,4634-40-42 triplet to asses the luminosity classes of the stars. Given that He\,{\sc ii} $\lambda$\,4686 is seen in strong absorption and N\,{\sc iii} $\lambda\lambda$\,4634-40-42 in weak absorption, both stars are of luminosity class V((f)).

We estimated the brightness ratio of HD\,165052 in the visible domain based on the ratio between the equivalent widths (EWs) of the spectral lines of the  secondary star and {\tt TLUSTY} spectra of similar effective temperatures. We used the H$\beta$, H$\gamma$, He\,{\sc i} $\lambda\lambda$\,4026, 4471, 5016, and He\,{\sc ii} $\lambda$\,4542 lines. The ratio $\text{EW}_{\text{\tt TLUSTY}}/\text{EW}_{\text{sec}} = (l_\text{P}+l_\text{S})/l_\text{S}$ amounts to $2.42\pm 0.39$ and $2.43\pm0.37$ when \texttt{TLUSTY} spectra of $T_\text{eff}$ of 37\,500 and 35\,000\,K, respectively, are used. Both results give us $l_\text{S}/l_\text{P} = 0.70\pm0.19$.

\citet{bailer21} derived a distance of $1212.7^{+42.8}_{-31.5}$\,pc from the parallax of $\varpi = 0.7893\pm 0.0297$\,mas quoted by the  {\it Gaia} early data release 3 \citep[EDR3,][]{EDR3}. This leads to a distance modulus of $10.42^{+0.08}_{-0.06}$ for the binary system. Mean $V$ and $B$ magnitudes of $6.87$ and $6.98$, respectively, are reported by \citet{Zacharias13}, for which we estimated errors of 0.01. We adopted a value of $-0.27\pm0.01$ for the intrinsic colour index $(B-V)_0$ of an O6.5-O7\,V star \citep{MP} and assumed the reddening factor in the $V$-band $R_V$ equal to $3.15\pm0.06$ for NGC\,6530 \citep{topasna20}. In this way, we obtained an absolute magnitude in the $V$-band of the binary system $M_V=-4.75^{+0.10}_{-0.08}$. Using the brightness ratio, we then derived individual absolute magnitudes $M_{V,\text{P}} =-4.17\pm0.15$ and $M_{V,\text{S}} =-3.79^{+0.20}_{-0.19}$ for the primary and secondary stars, respectively.
 
From the comparison with the magnitudes reported by \citet{MP}, we observe that the primary star is slightly less luminous than a `typical' O6.5\,V star while the secondary star is fainter than expected for O7-O7.5\,V type stars.

\subsection{Projected rotational velocities}\label{subsect:vsini}
\begin{figure*}
\centering
\includegraphics[width=0.49\linewidth, trim=2cm 0cm 3cm 6.5cm, clip=true]{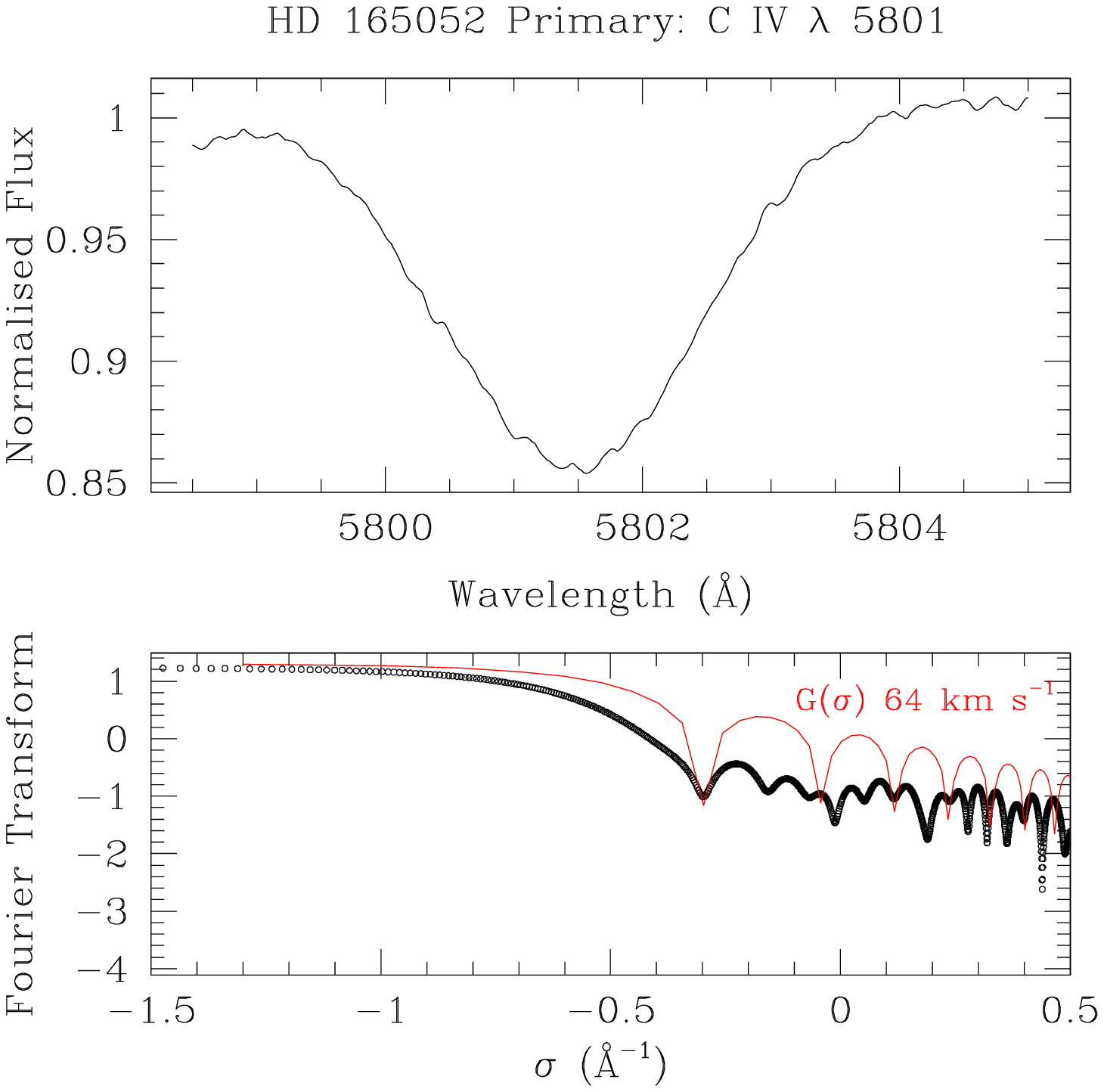}
\includegraphics[width=0.49\linewidth, trim=2cm 0cm 3cm 6.5cm, clip=true]{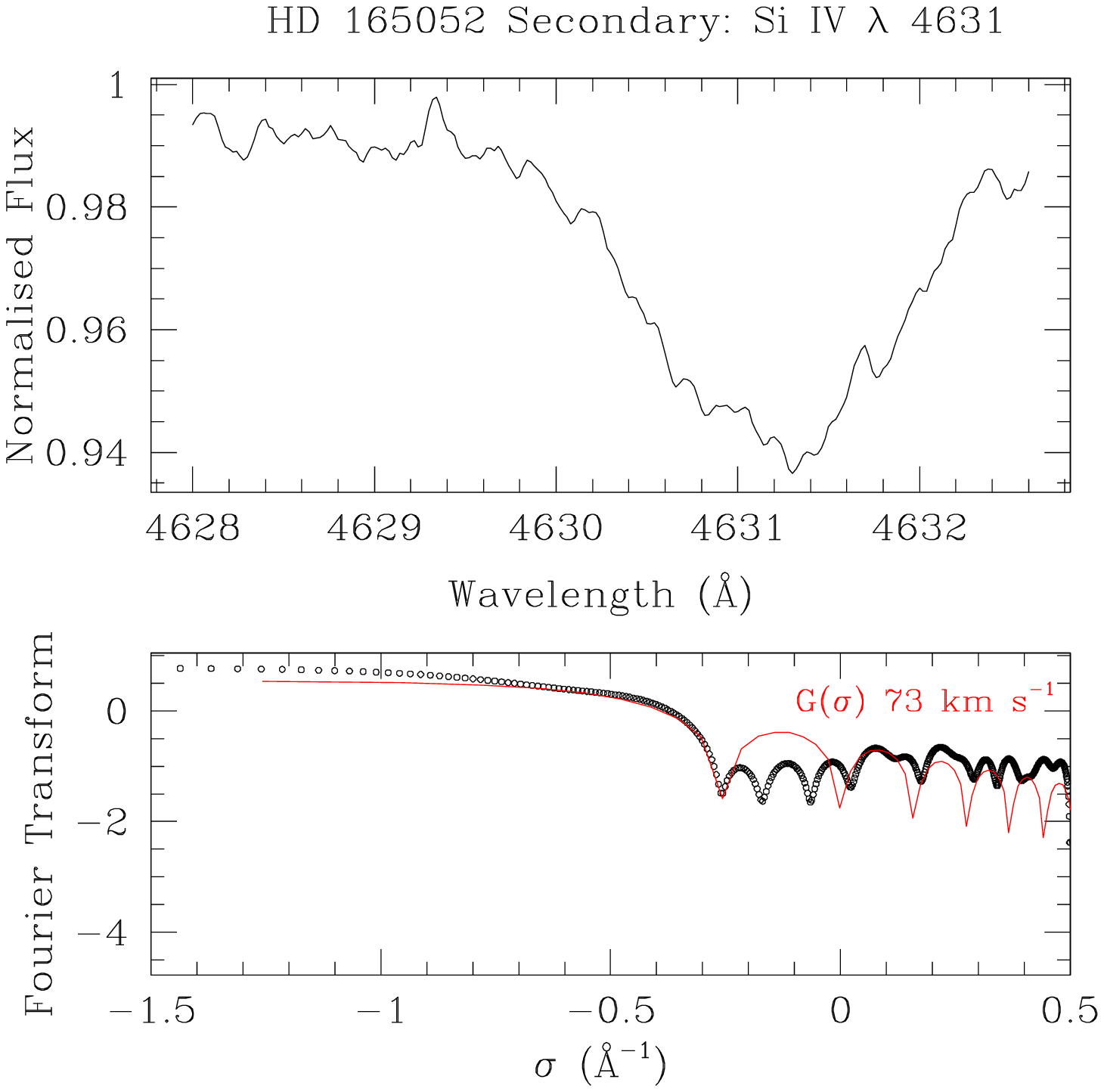}
\caption{Fourier transforms of primary and secondary lines. \textit{Top row:} Line profiles of the separated spectra of HD\,165052 obtained after application of the brightness ratio for the primary (C\,{\sc iv} $\lambda$\,5801 line, \textit{left panel}) and secondary (Si\,{\sc iv} $\lambda$\,4631 line, \textit{right panel}) stars. \textit{Bottom row:} Fourier transform of those lines (in black) and best-match rotational profile (in red) for the primary (\textit{left panel}) and secondary (\textit{right panel}) stars.}
 \label{fig:vsini}
\end{figure*}

We used the Fourier transform method \citep{Simon-Diaz,Gray08} and proceeded as in \citet{rosu22a} to derive the projected rotational velocities of both stars. The results are summarised in Table\,\ref{vsiniTable}, and the Fourier transforms of the C\,\textsc{iv} $\lambda$\,5801 line for the primary star and of the Si\,\textsc{iv} $\lambda$\,4631 line for the secondary star are illustrated in Fig.\,\ref{fig:vsini}. 
The results presented in Table\,\ref{vsiniTable} show that the mean $v \sin i_\text{rot}$ computed on metallic lines alone or on all the lines agree very well. We adopted a mean $v \sin i_\text{rot}$ of $67\pm8$ km\,s$^{-1}$ for the primary star and of $62\pm7$ km\,s$^{-1}$ for the secondary star. Our values are compatible, within the error bars, with those derived by \citet{morrison78}, \citet{stickland97}, \citet{linder07}, and \citet{FER} but are slightly lower than those quoted by \citet{howarth97}.

\begin{table}
\caption{Best-fit projected rotational velocities as derived from the disentangled spectra of HD\,165052 and comparison with results coming from the literature.}
\centering
\begin{tabular}{l l l}
\hline\hline
\vspace{-3mm}\\
Line & \multicolumn{2}{c}{$v\sin i_{\text{rot}}$ (km\,s$^{-1}$)} \\
& Primary & Secondary \\
\hline
\vspace{-3mm}\\
Si\,{\sc iv} $\lambda$\,4089 & 75 & 56 \\ 
Si\,{\sc iv} $\lambda$\,4116 & ... & 59 \\ 
Si\,{\sc iv} $\lambda$\,4212 & 66 & ... \\ 
Si\,{\sc iv} $\lambda$\,4631 & 55 & 73 \\ 
O\,{\sc iii} $\lambda$\,5592 & 73 & 59 \\
C\,{\sc iv} $\lambda$\,5801 & 64 & ... \\
C\,{\sc iv} $\lambda$\,5812 & ... & 61 \\
He\,{\sc i} $\lambda$\,3820 & 79 & ... \\ 
He\,{\sc i} $\lambda$\,4120 & ... & 73 \\
 He\,{\sc i} $\lambda$\,4143 & 74 & ... \\ 
He\,{\sc i} $\lambda$\,4387 & 73 & 77 \\ 
He\,{\sc i} $\lambda$\,4471 & ... & 73 \\ 
He\,{\sc i} $\lambda$\,4922 & ... & 57 \\ 
He\,{\sc i} $\lambda$\,5016 & 73 & ... \\
He\,{\sc i} $\lambda$\,5876 & ... & 62 \\
\hline
\vspace{-3mm}\\
Mean (metallic lines) & $67\pm8$ &$62\pm7$ \\ 
Mean (He\,{\sc i} lines) & $75\pm3$ &$68\pm8$ \\ 
Mean (all lines) & $70\pm7$ & $65\pm8$ \\
\hline
\vspace{-3mm}\\
\citet{stickland97} & $85 \pm 8$ & $80 \pm 6$\\
\citet{howarth97} & 91 & 78 \\
\citet{linder07} & $73\pm 7$ & $80 \pm 7$\\
\citet{FER} & $71 \pm 5$ & $66 \pm 5$\\
\hline
\end{tabular}
\begin{tablenotes}
\item The values quoted by \citet{stickland97} and \citet{howarth97} were obtained by cross-correlation techniques applied to \textit{IUE} spectra. The values quoted by \citet{linder07} were obtained by applying the Fourier method \citep[see][]{Simon-Diaz, gray05} to the profiles of their disentangled He\,{\sc i} $\lambda$\,4471, He\,{\sc ii} $\lambda$\,4542, and H$\beta$ line profiles. The values quoted by \citet{FER} were obtained using their empirical regressions between the full widths at half-maximum of intensity and $v \sin i_\text{rot}$ values derived based on the convolution of the spectrum of $\tau$\,Sco with rotation-line profiles.
\end{tablenotes}
\label{vsiniTable}
\end{table}

\subsection{Model atmosphere fitting}\label{subsect:cmfgen}
We analysed the reconstructed spectra of the binary components by means of the non-LTE model atmosphere code {\tt CMFGEN}  \citep{Hillier} to constrain the fundamental properties of the stars. We broadened the {\tt CMFGEN} spectra by the projected rotational velocities determined in Sect.\,\ref{subsect:vsini} and adjusted the stellar and wind parameters following the procedure outlined by \citet{Martins11}.

Given that the surface gravity derived in the $\texttt{CMFGEN}$ adjustment is underestimated for a binary star \citep{palate12}, we decided to fix the value of the surface gravity $\log g_\text{spectro}$ to 3.92\,cgs for both stars following Tables 1 and 4 in \citet{Martins05}. We fixed the microturbulence value at the level of the photosphere, $v_\text{micro}^\text{min}$, to 15\,km\,s$^{-1}$. The clumping parameters of the wind were fixed: the volume filling factor $f_1$ was set to 0.1, and the $f_2$ parameter controlling the onset of clumping was set to 100\,km\,s$^{-1}$, as in \citet{rosu20b}. Likewise, the $\beta$ parameter of the velocity law was fixed to 1.0 as suggested by \citet{muijres12} for O6.5-O7\,V type stars. The wind terminal velocity $v_\infty$ was fixed to 2335\,km\,s$^{-1}$ for both stars, as derived from the combined \textit{IUE} spectra by \citet{howarth97}.

We adjusted the macroturbulence velocity $v_\text{macro}$ on the wings of the O\,{\sc iii} $\lambda$\,5592 and Balmer lines and derived values of $120\pm 20$\,km\,s$^{-1}$ and $65\pm10$\,km\,s$^{-1}$ for the primary and secondary stars, respectively.

\begin{figure*}
\centering
\includegraphics[clip=true,trim=15 135 35 25,width=\linewidth]{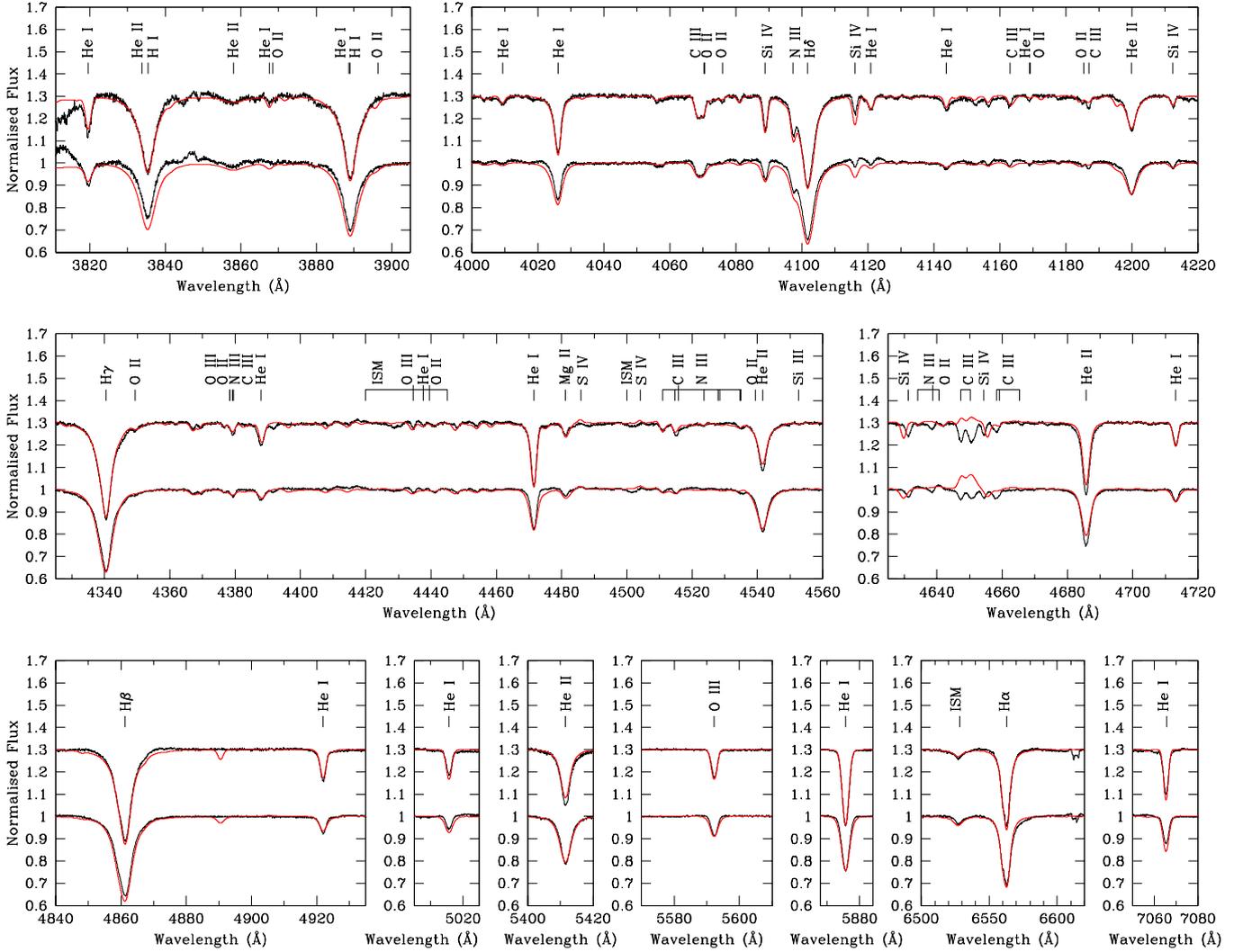}
\caption{Normalised disentangled spectra (in black) of the primary and secondary stars of HD\,165052 (the spectrum of the secondary star is shifted by +0.3 in the $y$-axis for clarity) together with the respective best-fit {\tt CMFGEN} model atmosphere (in red).}
\label{Fig:CMFGEN}
\end{figure*}

The stellar and wind parameters of the best-fit {\tt CMFGEN} model atmosphere are summarised in Table\,\ref{Table:CMFGEN}. The normalised disentangled spectra of HD\,165052 are illustrated in Fig.\,\ref{Fig:CMFGEN} along with the best-fit {\tt CMFGEN} adjustments.

We derived the effective temperature based on the adjustment of the He\,{\sc i} $\lambda\lambda$\,4121, 4471, 4713, 4922, 5016, 5874, 7065 and He\,{\sc ii} $\lambda\lambda$\,4200, 4542, 5412 lines. This was clearly a compromise as we could not find a solution that perfectly fits all helium lines simultaneously. We discarded the He\,{\sc i} $\lambda$\,4026 line because of its blend with the weak but non-zero He\,{\sc ii} $\lambda$\,4026 line. We derived effective temperatures $T_\text{eff,1} = 37\,500 \pm 1000$\,K and $T_\text{eff,2} = 36\,000 \pm 1000$\,K for the primary and secondary stars, respectively. 

The mass-loss rate was derived based on the adjustment of the H$\alpha$ and He\,{\sc ii} $\lambda$\,4686 lines. We derived values of $(1.5\pm 0.5)\times 10^{-7}\,M_\odot$\,yr$^{-1}$ and $(9.0\pm 1.0)\times 10^{-8}\,M_\odot$\,yr$^{-1}$ for the primary and secondary stars, respectively. 

We set the surface chemical abundances of all elements, including helium, but excluding carbon and oxygen, to solar as taken from \citet{asplund09}. We derived the oxygen abundance based on the O\,{\sc iii} $\lambda$\,5592 line as it is the sole oxygen line free of blends. We derived sub-solar oxygen abundances in number O/H of $(3.0\pm 0.2)\times 10^{-4}$ and $(4.0\pm 0.2)\times 10^{-4}$ for the primary and secondary stars, respectively. For both stars, with these oxygen abundances, the weak O\,{\sc iii} $\lambda\lambda$\,4368, 4448, 4454, 4458 lines are well-reproduced, while the O\,{\sc iii} $\lambda$\,5508 line is slightly overestimated. We derived the carbon abundance based on the C\,{\sc iii} $\lambda$\,4070 line as it is the sole reliable carbon line free of blends. Indeed, as stated by \citet{Martins12}, the C\,{\sc iii} $\lambda\lambda$\,4647-51 blend and the C\,{\sc iii} $\lambda$\,5696 line are known to be problematic because their formation processes are controlled by a number of far-UV lines, hence their strength and nature critically depend upon subtle details of the stellar atmosphere model. The C\,{\sc iii} $\lambda$\,4379 line is heavily blended with the N\,{\sc iii} $\lambda$\,4379 and therefore useless, while the C\,{\sc iii} $\lambda$\,4388 line is not significantly affected by a change in carbon abundance. We also know the C\,{\sc iv} $\lambda$\,5801, 5812 lines to be problematic and rarely correctly reproduced \citep{rosu20b}. We derived a sub-solar carbon abundances in number C/H of $(2.0\pm 0.2)\times 10^{-4}$ for both stars.  
Finally, we set the nitrogen abundance to solar as taken from \citet{asplund09} as we could not adjust the nitrogen abundance based on the nitrogen lines. Indeed, N\,{\sc iii} $\lambda$\,4379 is heavily blended with C\,{\sc iii} $\lambda$\,4379, the N\,{\sc iii} $\lambda\lambda$\,4510-4540 blend is not significantly affected by a change in nitrogen abundance, and the \texttt{CMFGEN} spectra display the N\,{\sc iii} $\lambda\lambda$\,4634-4640 complex in emission. We tested a nitrogen abundance 4.83 (respectively 3.35) times solar for the primary (respectively secondary) star -- such that, combined with the depleted C and O abundances, the CNO abundance is solar -- and observed that the nitrogen lines in the \texttt{CMFGEN} spectra were much deeper than in the observational spectra. This results suggests that the initial metallicity of the stars is (at least slightly) sub-solar.

\begin{table}
\caption{Stellar and wind parameters of the best-fit {\tt CMFGEN} model atmosphere derived from the separated spectra of HD\,165052.}
\centering
\begin{tabular}{l l l}
\hline\hline
\vspace{-3mm}\\
Parameter & \multicolumn{2}{c}{Value} \\
& Primary & Secondary \\
\hline
\vspace{-3mm}\\
$T_\text{eff}$ (K) & $37\,500\pm1000$ & $36\,000\pm1000$  \\ 
$\log g_\text{spectro}$ (cgs) & $3.92$ (fixed) & $3.92$ (fixed) \\
$v_\text{macro}~(\text{km\,s}^{-1})$ & $120\pm20$ & $65\pm10$ \\ 
$v_\text{micro}^\text{min}~(\text{km\,s}^{-1})$ & $15$ (fixed) & $15$ (fixed)\\ [0.1cm]
$\dot{M}~(M_\odot\,\text{yr}^{-1})$  & $(1.5\pm0.5)\times10^{-7}$ &$(9.0\pm1.0)\times 10^{-8}$ \\
$\dot{M}_\text{uncl.}~(M_\odot\,\text{yr}^{-1})^a$  & $(4.7\pm 1.6)\times10^{-7}$ & $(2.8\pm 0.3)\times10^{-7}$ \\
$v_\infty~(\text{km\,s}^{-1})$ & 2335 (fixed)  & 2335 (fixed) \\
$f_1$ & 0.1 (fixed) & 0.1 (fixed)  \\ 
$f_2~(\text{km\,s}^{-1})$ & 100 (fixed) & 100 (fixed)\\
$\beta$ & 1.1 (fixed) & 1.1  (fixed) \\
He/H (nb) & 0.0851$^b$ (fixed) & 0.0851$^b$ (fixed) \\
C/H (nb) & $(2.0\pm 0.2)\times 10^{-4}$ & $(2.0\pm 0.2)\times 10^{-4}$  \\
N/H (nb) & $6.76 \times 10^{-5}$ (fixed$^b$) & $6.76 \times 10^{-5}$ (fixed$^b$) \\
O/H (nb) & $(3.0\pm 0.2)\times 10^{-4}$ & $(4.0\pm 0.2)\times 10^{-4}$\\
\vspace*{-3mm}\\
\hline
\end{tabular}
\begin{tablenotes}
\item $^a$$\dot{M}_\text{uncl.}=\dot{M}/\sqrt{f_1}$ is the unclumped mass-loss rate. 
\item $^b$Value fixed to the solar chemical abundance \citep{asplund09}.
\end{tablenotes}
\label{Table:CMFGEN}
\end{table} 

We computed the bolometric magnitudes of the stars assuming the bolometric correction depends only on the effective temperature through the relation 
\begin{equation}
\text{BC} = -6.89\log(T_\text{eff})+28.07
\end{equation}
\citep{MP} and got $M_\text{bol,P} = -7.61\pm 0.17$ and $M_\text{bol,S} = -7.11\pm 0.21$, which then converted into bolometric luminosities give us $L_\text{bol,P} = 89\,000\pm 14\,000\,L_\odot$ and $L_\text{bol,S} = 56\,000 \pm 11\,000\,L_\odot$. Combined with the effective temperatures derived from the \texttt{CMFGEN} analysis, we inferred spectroscopic radii $R_\text{spectro,P}=7.1 \pm 0.7\,R_\odot$ and $R_\text{spectro,S}=6.1 \pm 0.7\,R_\odot$. The surface gravities, corrected for both the centrifugal force and radiation pressure, amount to $3.99\pm 0.01$ and $3.98^{+0.02}_{-0.01}$ for the primary and secondary stars, respectively. Spectroscopic masses $M_\text{spectro,P}=17.7^{+3.5}_{-3.4}\,M_\odot$ and $M_\text{spectro,S}=12.9\pm2.9\,M_\odot$ were then inferred.

\section{Radial velocity analysis}\label{sect:omegadot}
Our total set of RV measurements consists of our 71 primary and secondary RVs determined as part of the disentangling process (see Sect.\,\ref{subsect:spectraldisentangling}) complemented by 94 primary and secondary RVs coming from the literature. Twelve RVs come from \citet{morrison78} but we adopted the corrected values by \citet{stickland97}. The latter authors do not report uncertainties on their measurements, and we therefore assumed symmetric uncertainties of 15\,km\,s$^{-1}$, as representative of the O-C. Fifteen RVs come from \citet{stickland97}, and we assumed symmetric uncertainties on these RVs of 10\,km\,s$^{-1}$ as representative of the O-C. Thirty RVs come from \citet{arias02} and (J. I. Arias, private communication, 2021), for which we adopted the uncertainties quoted by the authors. The remaining 37 RVs come from \citet{FER} that we recalculated here (see Appendix\,\ref{appendix:RVs_ferrero}). We ended up, in total, with a series of 165 RVs spanning about 46 years. For the RVs derived from the spectral disentangling, we adopted formal errors of 3\,km\,s$^{-1}$, as the small errors derived as part of the disentangling method would bias the adjustment given our high number of RVs compared to those coming from the literature. We then processed in the adjustment of the RVs along three different avenues.

\begin{figure*}
\centering
\hspace{-6.93cm}\includegraphics[clip=true,trim=0cm 2.3cm 0.8cm 7cm, width=0.615\linewidth]{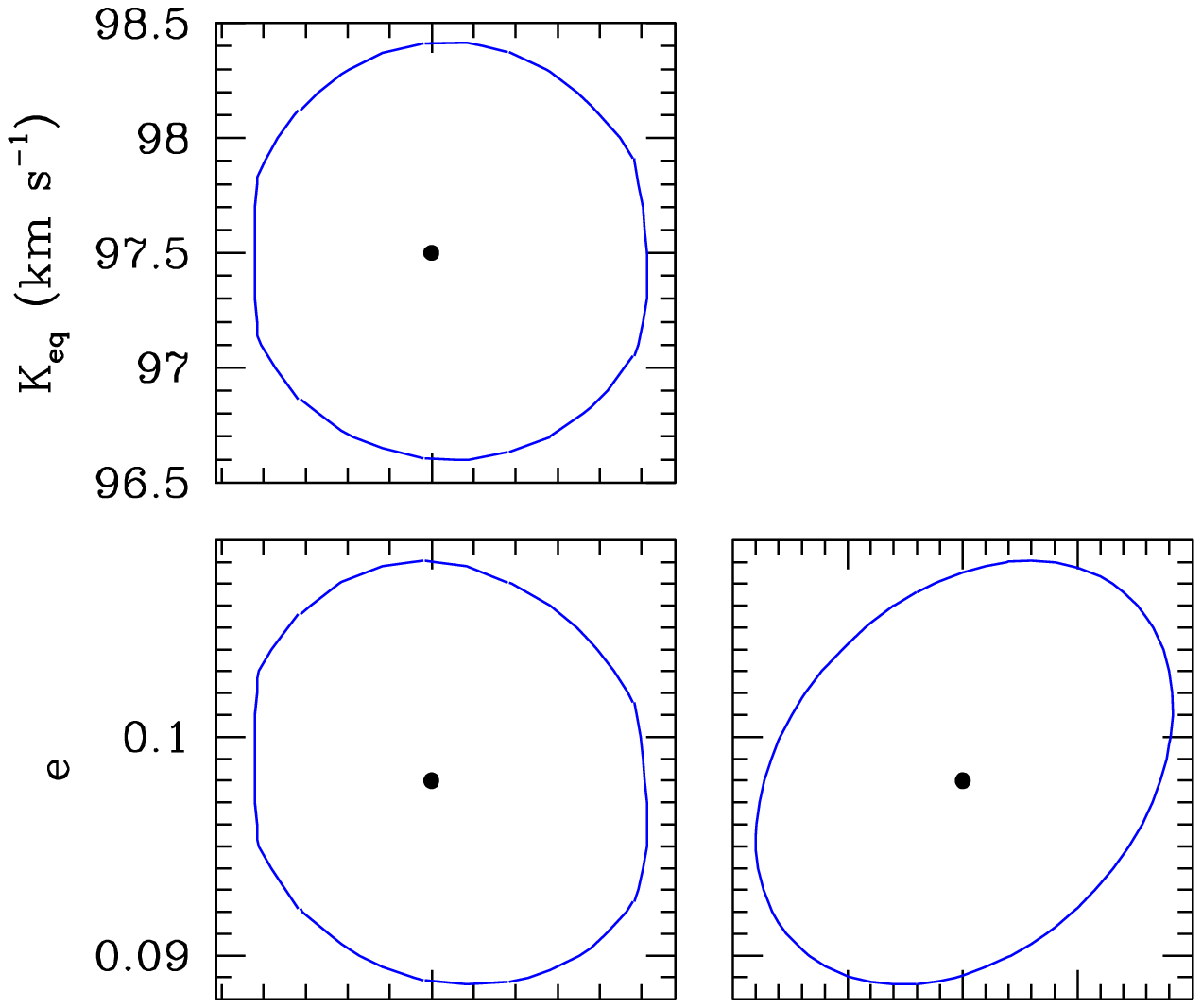}\\
\vspace{-0.7cm}
\includegraphics[clip=true,trim=0cm 0.98cm 0.8cm 1.4cm, width=0.615\linewidth]{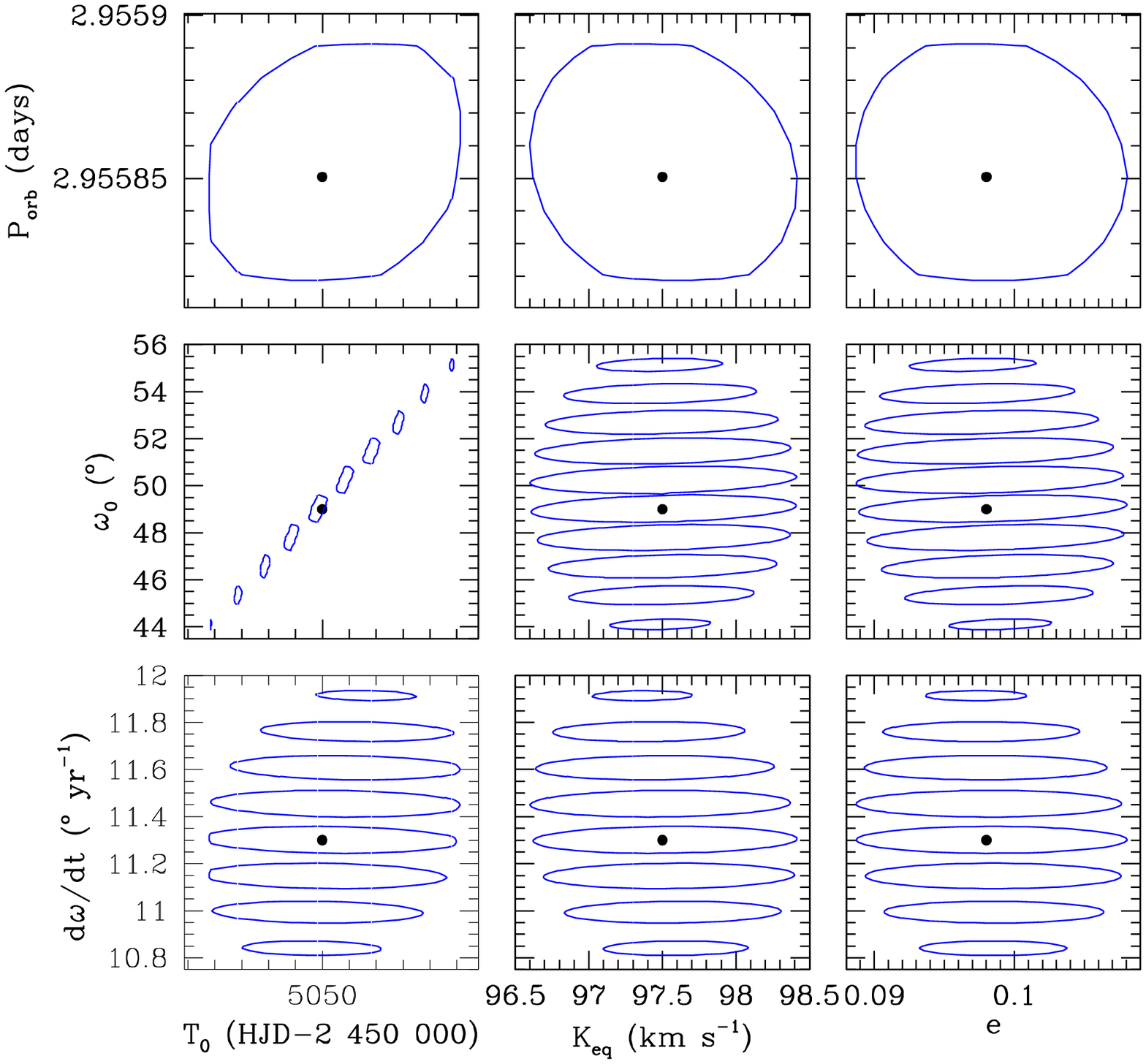}
\includegraphics[clip=true,trim=2.8cm 0.98cm 5.57cm 0cm, width=0.375\linewidth]{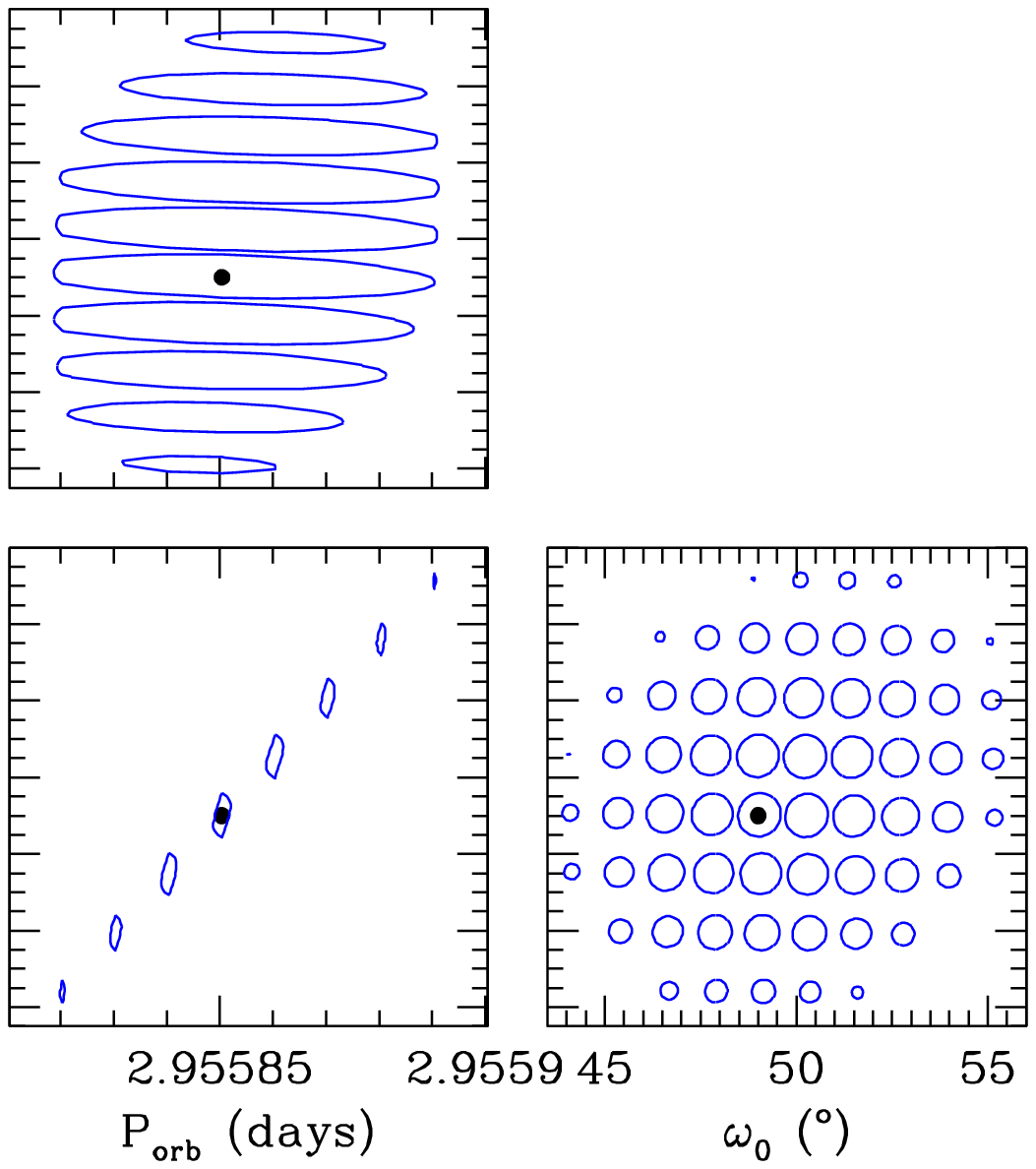}
\caption{Confidence contours for the best-fit parameters obtained from the adjustment of the equivalent RV data of HD\,165052. The best-fit solution is shown in each panel by the black filled dot. The corresponding $1\sigma$ confidence level is shown by the blue contour. We note the “zebra-like” strips for $\omega_0$ and $\dot\omega$ are computational artefacts.}
\label{fig:contours_RVs_eq}
\end{figure*}

\begin{figure*}
\centering
\includegraphics[clip=true, trim=5 175 25 0,width=0.8\linewidth]{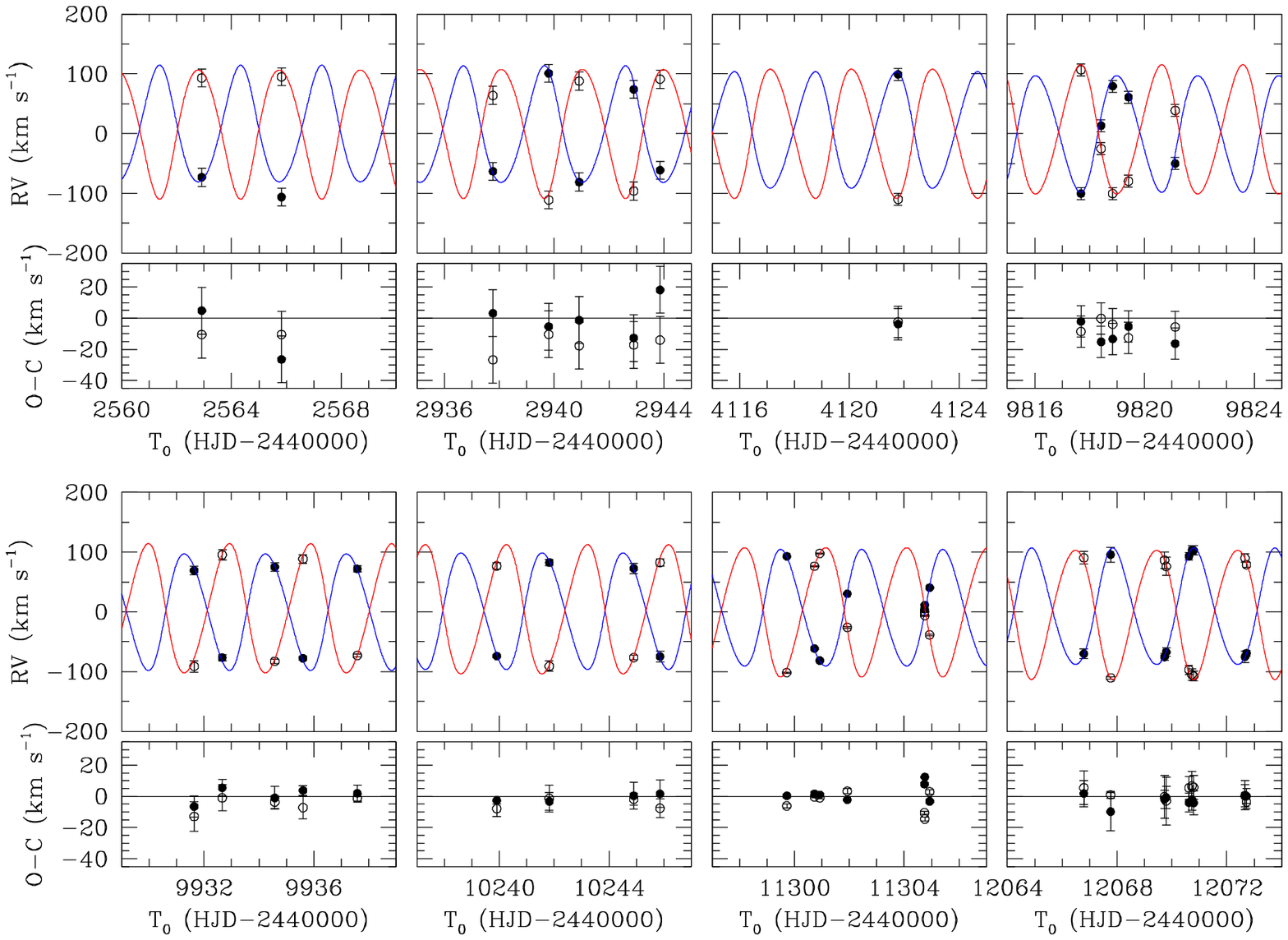}
\includegraphics[clip=true, trim=5 175 25 0,width=0.8\linewidth]{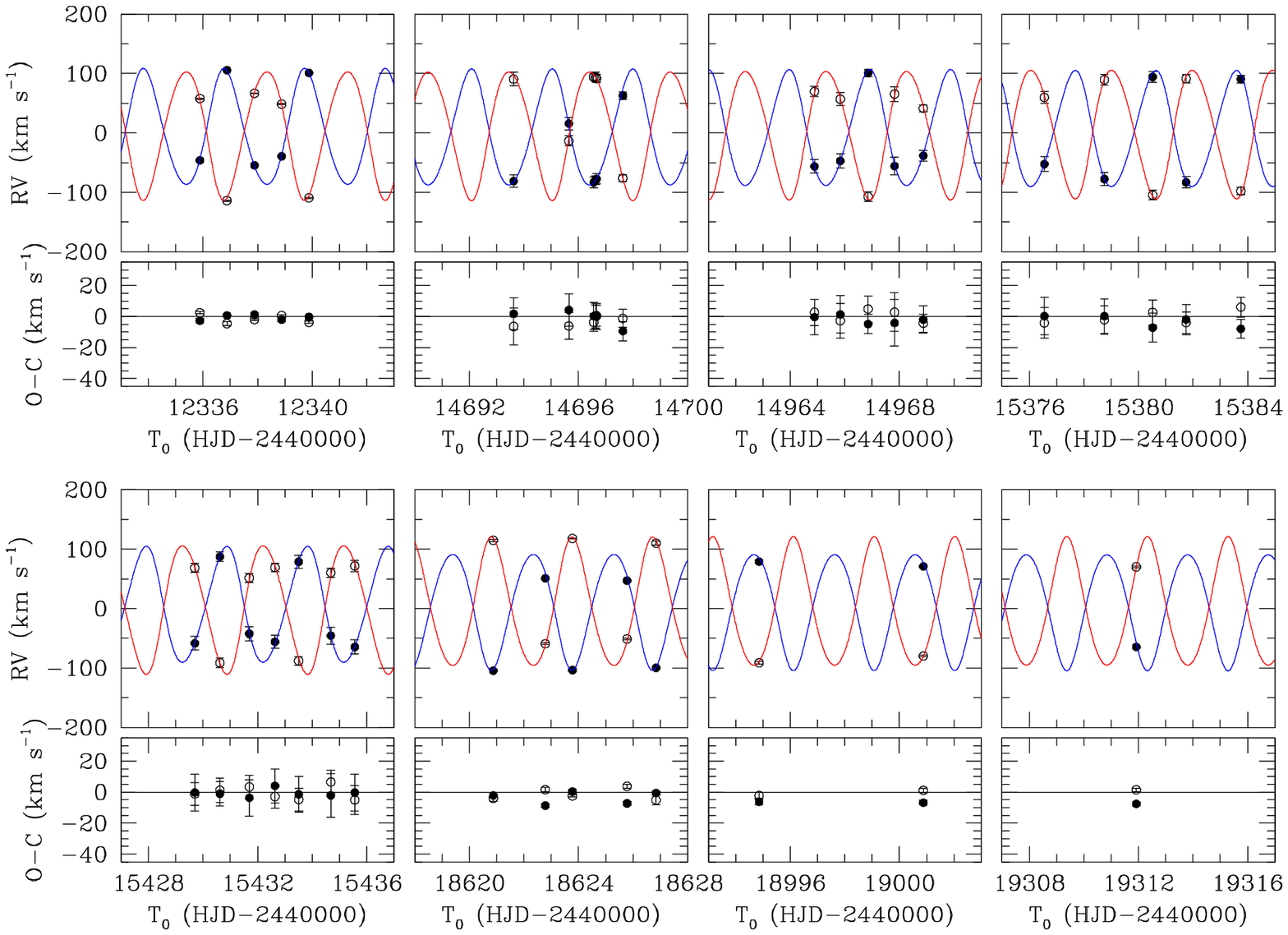}
\caption{Comparison between the measured RVs of the primary (filled dots) and secondary (open dots) of HD\,165052 with the orbital solution from the equivalent RVs (see last column in Table\,\ref{table:RVs_fit}). The blue and red lines represent the fitted RV curve of the primary and secondary stars, respectively.  The top panels correspond to data from \citet[][\textit{two left ones}]{morrison78} and \citet[][\textit{two right ones}]{stickland97}. The panels on the second row correspond to data from \citet[][\textit{the first, second, and fourth ones}]{arias02} and to RVs derived in this paper (\textit{third one}). The panels on the third row correspond to RVs derived in this paper (\textit{first one}) and data from \citet[][\textit{last three ones}]{FER}.  The last row corresponds to data from \citet[][\textit{first one}]{FER} and to RVs derived in this work (\textit{last three ones}).}
\label{fitRV}
\end{figure*}

\begin{table*}
\caption{Best-fit orbital parameters of HD\,165052 obtained from the adjustment of the RV data.}
\label{table:RVs_fit}
\centering
\begin{tabular}{l l l l}
\hline\hline
\vspace{-3mm}\\
Parameter & Primary RVs & Secondary RVs & Equivalent RVs \\
\hline
\vspace*{-3mm}\\
$e$ & $0.093^{+0.009}_{-0.008}$ & $0.096\pm0.007$ & $0.098\pm 0.010$ \\ 
\vspace*{-3mm}\\
$\dot{\omega}$ ($^{\circ}$\,yr$^{-1}$) & $12.02^{+0.63}_{-0.48}$ & $10.72^{+0.50}_{-0.34}$ & $11.30^{+0.64}_{-0.49}$ \\
\vspace*{-3mm}\\
$P_{\rm orb}$\,(d) &  $2.95590^{+0.00004}_{-0.00003}$ & $2.95581\pm 0.00003$ & $2.95585^{+0.00004}_{-0.00003}$\\
\vspace*{-3mm}\\
$\omega_0$ ($^\circ$)  & $52.8^{+5.2}_{-6.3}$ & $50.1^{+3.9}_{-5.1}$ & $49.0^{+6.4}_{-5.2}$\\
\vspace*{-3mm}\\
$T_0$ (HJD) & $2\,455\,050.03^{+0.04}_{-0.05}$ & $2\,455\,050.01^{+0.04}_{-0.05}$ & $2\,455\,050.00^{+0.05}_{-0.04}$\\
\vspace*{-3mm}\\
$K_{\rm P}$\,(km\,s$^{-1}$) & $97.3\pm0.8$ & $96.8\pm1.2$ & $97.5\pm0.9$\\
\vspace*{-3mm}\\
$K_{\rm S}$\,(km\,s$^{-1}$) & $107.9\pm1.4$ & $107.4^{+0.7}_{-0.8}$ & $108.1\pm1.5$\\
\vspace*{-3mm}\\
$K_{\rm eq}$\,(km\,s$^{-1}$) &  ... & ... & $97.5 \pm 0.9$\\
\vspace*{-3mm}\\
$q = M_\text{S}/M_\text{P}$   & $0.902\pm 0.009$& $0.902\pm 0.009$& $0.902\pm 0.009$ \\
\vspace*{-3mm}\\
$a_{\rm P}\,\sin{i}$\,($R_{\odot}$) & $5.66\pm0.04$ & $5.63\pm0.05$ &$5.66\pm0.05$ \\
\vspace*{-3mm}\\
$a_{\rm S}\,\sin{i}$\,($R_{\odot}$) & $6.27\pm0.06$ & $6.24^{+0.04}_{-0.05}$ & $6.28\pm0.06$\\
\vspace*{-3mm}\\
$M_{\rm P}\,\sin^3{i}$\,($M_{\odot}$) & $1.37 \pm 0.04$ & $1.35\pm0.04$ & $1.38\pm0.05$\\
\vspace*{-3mm}\\
$M_{\rm S}\,\sin^3{i}$\,($M_{\odot}$) & $1.24\pm0.04$  & $1.22^{+0.03}_{-0.04}$ & $1.24\pm0.04$\\
\vspace*{-3mm}\\
$\chi^2_\nu$ & 1.465 & 2.155 & 0.971\\
\vspace*{-3mm}\\
\hline
\end{tabular}
\begin{tablenotes}
\item In the case of the primary (respectively secondary) RVs, we used the mass ratio derived in Sect.\,\ref{sect:omegadot} to convert $K_\text{P}$ (respectively $K_\text{S}$) into a value for $K_\text{S}$ (respectively $K_\text{P}$) and compute the minimum masses but we did not adjust the secondary (respectively primary) RVs directly.
\end{tablenotes}
\end{table*}

First, for each time of observation $t$, we adjusted the primary RV data with the following relation 
\begin{equation}
\label{eqn:RVp}
\text{RV}_\text{P}(t) = \gamma_\text{P} + K_\text{P} [\cos(\phi(t)+\omega(t)) +e\cos\omega(t)],
\end{equation}
where $\gamma_\text{P}$, $K_\text{P}$, $e$, and $\omega$ are the primary apparent systemic velocity, the semi-amplitude of the primary RV curve, the eccentricity, and the argument of periastron of the primary orbit, respectively. The true anomaly $\phi$ is inferred from the eccentric anomaly, itself computed through Kepler's equation, which involves both $e$ and the anomalistic orbital period $P_\text{orb}$ of the system. We accounted for the apsidal motion through the variation of $\omega$ with time following the relation 
\begin{equation}
\omega(t) = \omega_0 + \dot\omega(t-T_0),
\end{equation}
where $\dot\omega$ is the apsidal motion rate and $\omega_0$ is the value of $\omega$ at the time of periastron passage $T_0$. Given that different spectral lines give potentially slightly different values of the apparent systemic velocities, and given that the different RVs were obtained based on different sets of lines, the systemic velocity of each subset of our total dataset was adjusted so as to minimise the sum of the residuals of the data about the curve given by Eq.\,\eqref{eqn:RVp}.

Second, we adjusted the secondary RV data with a relation similar to Eq.\,\eqref{eqn:RVp} corresponding to the secondary star, with $K_\text{S}$ and $\gamma_\text{S}$ having straightforward definitions.

Third, we made use of the linear relation 
\begin{equation}
\text{RV}_\text{P}(t)=-q\text{RV}_\text{S}(t)+B
\end{equation}
relating the primary and secondary RVs of an SB2 system, where $q=\frac{M_\text{S}}{M_\text{P}}$ is the mass ratio and $B=\gamma_\text{P}+q\gamma_\text{S}$, to convert the RVs of both stars into equivalent RVs of the primary star with
\begin{equation}
\text{RV}_\text{eq}(t)=\frac{\text{RV}_\text{P}-q\text{RV}_\text{S}+B}{2}.
\end{equation}
to which we associate corresponding $K_\text{eq}$ and $\gamma_\text{eq}$. We derived a value $q =0.902 \pm 0.009$ from the RVs coming from the disentangling process.

In the three cases, we scanned the 6-D parameter space in a systematic way to find the values of the free parameters ($P_{\rm orb}$, $e$, $T_0$, $\omega_0$, $\dot\omega$, and $K_{\rm P}$ or $K_{\rm S}$ or $K_{\rm eq}$) that provide the best fit to the whole set of corresponding RV data. The projections of the 6-D parameter space onto the 2-D planes is illustrated in Fig.\,\ref{fig:contours_RVs_eq} for the equivalent RVs. The corresponding orbital parameters are given in Table\,\ref{table:RVs_fit}. $a_\text{P}\sin i$ and $a_\text{S}\sin i$ stand for the minimum semi-major axis of the primary and secondary stars, respectively, $M_\text{P}\sin^3 i$ and $M_\text{S}\sin^3 i$ stand for the minimum mass of the primary and secondary stars, respectively, and $\chi^2_\nu$ is the reduced $\chi^2$. We note that in the case of the primary (respectively secondary) RVs, we used the mass ratio derived previously to convert $K_\text{P}$ (respectively $K_\text{S}$) into a value for $K_\text{S}$ (respectively $K_\text{P}$) and compute the associated minimum masses but we did not adjust the secondary (respectively primary) RVs directly. Figure\,\ref{fitRV} illustrates the best fit of the RV data with the equivalent RVs solution at 16 different epochs.

The eccentricities we obtained for the three solutions are compatible, within the error bars, with those coming from the literature (see Table\,\ref{table:intro}), except with the value of \citet{stickland97} as these authors explicitly assumed a zero eccentricity for the system. Our orbital periods are longer and not compatible, within the error bars, with those coming from the literature (we did not consider the one of \citet{morrison78}, see discussion in Sect.\,\ref{sect:intro}). Our value of $K_\text{P}$ is compatible, within the error bars, with those of \citet{stickland97}, \citet{linder07}, and \citet{FER}, but slightly larger than the one quoted by \citet{arias02}. Our value of $K_\text{S}$ is compatible with those of \citet{stickland97}, and \citet{FER} but slightly larger than the one quoted by \citet{arias02} and slightly smaller than the one quoted by \citet[see Table\,\ref{table:intro}]{linder07}. Our mass ratio is compatible, within the error bars, with those coming from the literature, except for the one quoted by \citet[see Table\,\ref{table:intro}]{linder07}. Finally, our apsidal motion rates are slightly smaller than the value of $(12.1\pm0.3)^\circ\,\text{yr}^{-1}$ reported by \citet{FER} but, in the cases of the primary and equivalent RVs solutions, still in agreement within the error bars. 

\section{Stellar structure and evolution tracks}
\label{sect:cles}
We computed stellar structure and evolution tracks with the Code Li\'egeois d'\'Evolution Stellaire\footnote{The \texttt{Cl\'es} code is developed and maintained by Richard Scuflaire at the STAR Institute at the University of Li\`ege.} \citep[\texttt{Cl\'es,}][see also \citet{rosu20a} for the main features of \texttt{Cl\'es} used in the present context]{scuflaire08}. The first goal of this theoretical analysis is to derive evolutionary masses for the stars and, hence, to put a constraint on the value of the inclination of the orbit through the confrontation with the minimum stellar masses derived in Sect.\,\ref{sect:omegadot}. The second goal of this analysis is to see how the theoretical apsidal motion rates derived for two stellar models of the same age compare to the observational apsidal motion rate. In the present analysis, we assumed that HD\,165052 has an age of $2.0\pm0.5$\,Myr (see Sect.\,\ref{sect:intro}).

The apsidal motion rate of a binary system, in the simple two-body case, is the sum of a Newtonian contribution (N) and a general relativistic correction (GR), which expressions were introduced by \citet{sterne} for the former one, and \citet{shakura} and \citet{gimenez85} for the latter one. We here adopt the same conventions and notations as in \cite{rosu20a} (see Sects.\,3 and 4, and Equations (17) to (20) and (5) to (9)).

\begin{figure*}
\includegraphics[clip=true, trim=120 100 350 180, width=\linewidth]{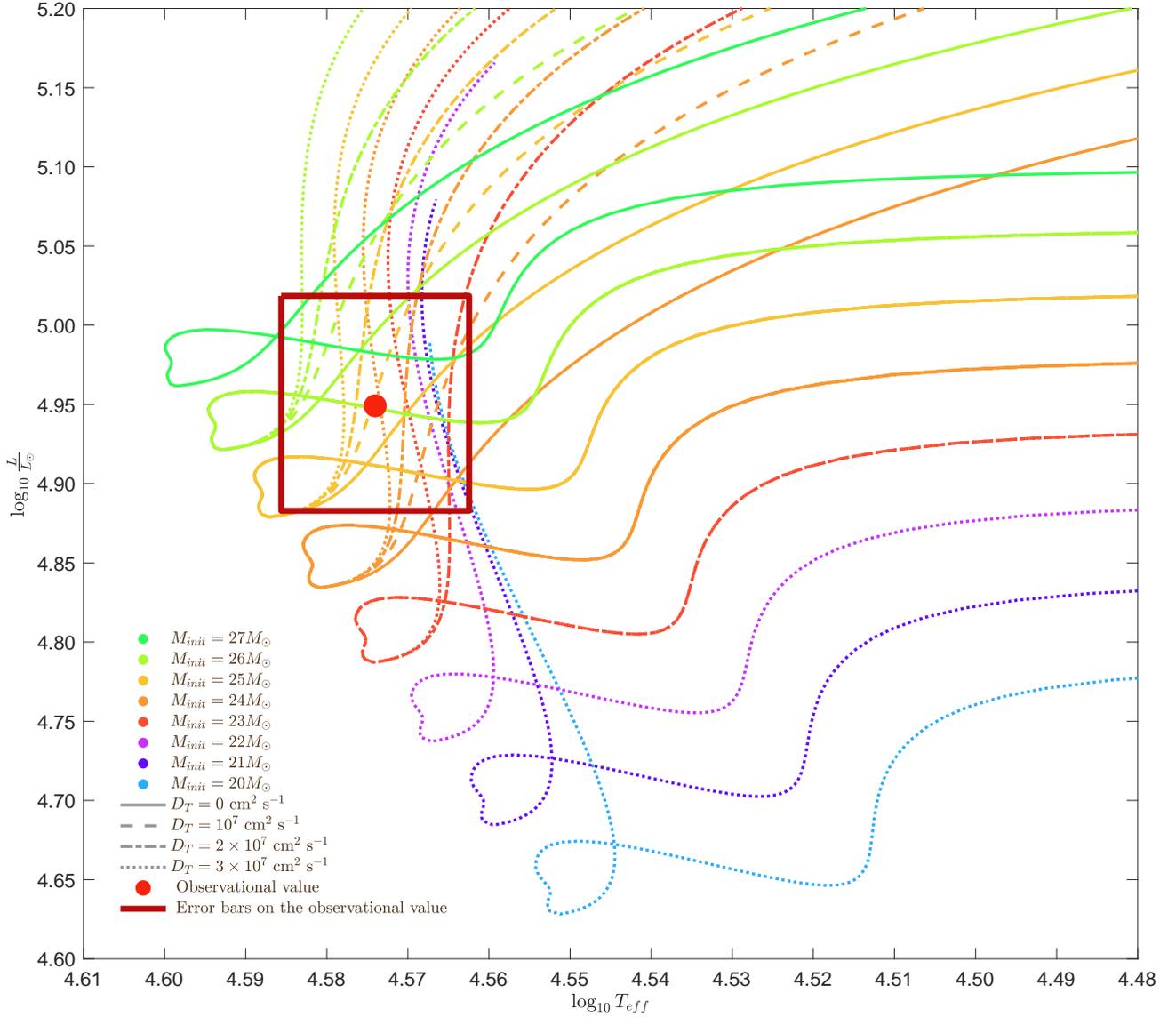}
\caption{Hertzsprung-Russell diagram: evolutionary tracks of \texttt{Cl\'es} models for the primary star of $M_\text{init}=20\,M_\odot$ (light blue), $21\,M_\odot$ (purple), $22\,M_\odot$ (plum), $23\,M_\odot$ (coral), $24\,M_\odot$ (orange), $25\,M_\odot$ (yellow), $26\,M_\odot$ (lime), and $27\,M_\odot$ (watergreen), and $D_T=0$\,cm$^2$\,s$^{-1}$ (solid line), $10^7$\,cm$^2$\,s$^{-1}$ (dashed line), $2\times10^7$\,cm$^2$\,s$^{-1}$ (dotted-dashed line), and $3\times10^7$\,cm$^2$\,s$^{-1}$ (dotted line). All models have $\xi=1$ and $\alpha_\text{ov}=0.30$. The observational value is represented by the red point, and its error bars are represented by the dark red rectangle.}
\label{fig:HRprim}
\end{figure*}

\begin{figure*}
\includegraphics[clip=true, trim=120 100 350 180, width=\linewidth]{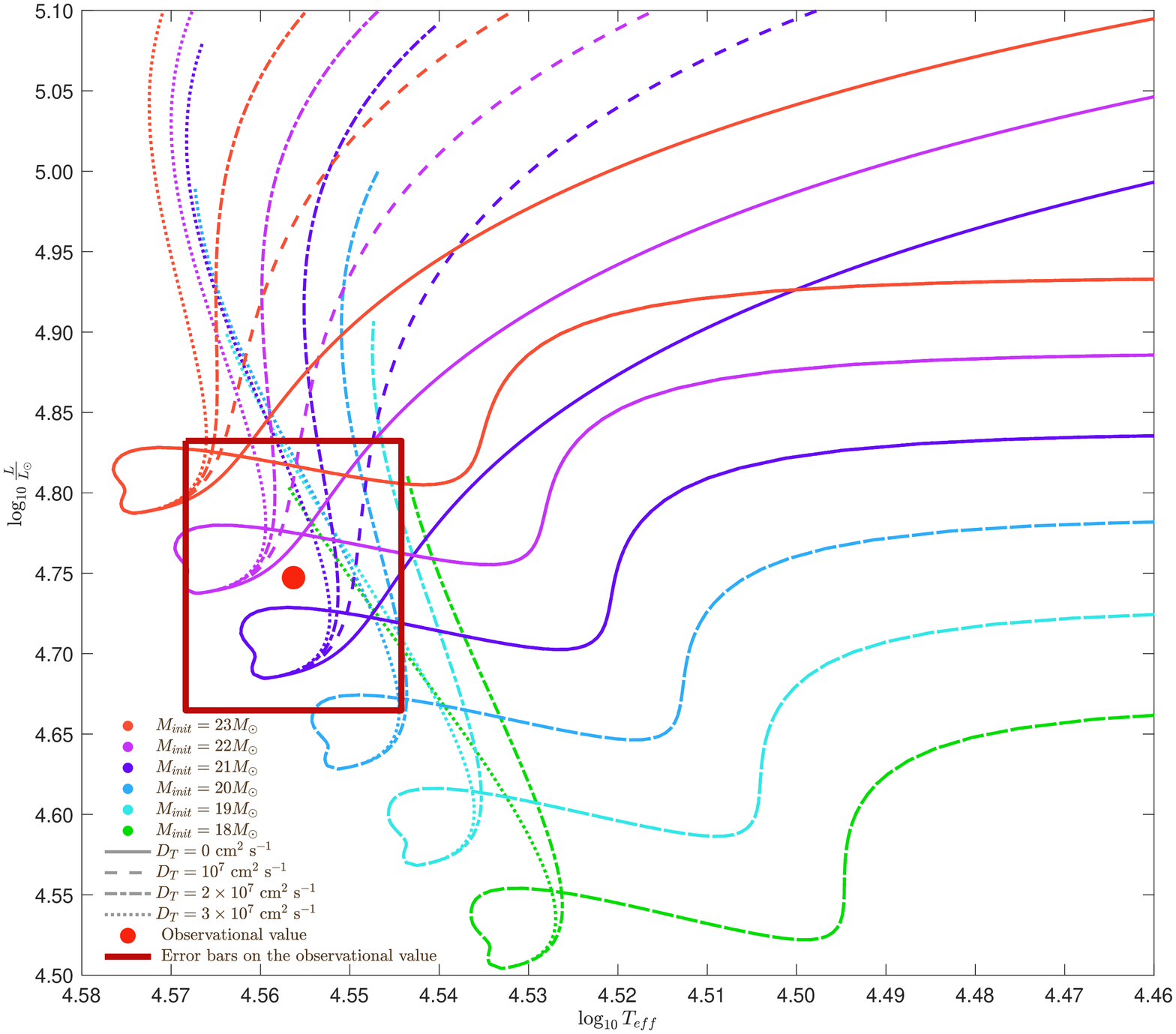}
\caption{Hertzsprung-Russell diagram: evolutionary tracks of \texttt{Cl\'es} models for the secondary star of $M_\text{init}=18\,M_\odot$ (green), $19\,M_\odot$ (turquoise), $20\,M_\odot$ (light blue), $21\,M_\odot$ (purple), $22\,M_\odot$ (plum), and $23\,M_\odot$ (coral), and $D_T=0$\,cm$^2$\,s$^{-1}$ (solid line), $10^7$\,cm$^2$\,s$^{-1}$ (dashed line), $2\times10^7$\,cm$^2$\,s$^{-1}$ (dotted-dashed line), and $3\times10^7$\,cm$^2$\,s$^{-1}$ (dotted line). All models have $\xi=1$ and $\alpha_\text{ov}=0.30$. The observational value is represented by the red point, and its error bars are represented by the dark red rectangle.}
\label{fig:HRsec}
\end{figure*}

We built a grid of stellar evolution tracks having an initial mass $M_\text{init}$ ranging from 18 to $27\,M_\odot$ and a turbulent diffusion coefficient $D_T$ of 0, $10^7$, $2\times10^7$, and $3\times10^7$\,cm$^2$\,s$^{-1}$, and for which we adopted a standard mass-loss scaling factor $\xi=1$ and an overshooting parameter $\alpha_\text{ov}=0.30$ \citep[see][for a description of these parameters and a discussion of their impact and their standard values]{rosu20a, rosu22a, rosu22b}. We present in the Hertzsprung-Russell diagrams in Figs.\,\ref{fig:HRprim} and \ref{fig:HRsec} the evolutionary tracks for the primary and secondary stars, respectively. For better visibility in the diagrams, we decided to only present the tracks that cross the observational boxes defined by the observational effective temperatures and bolometric luminosities of the stars. Among these selected models, we discarded those having an age lower than 1.5\,Myr or higher than 2.5\,Myr when crossing the observational box in the Hertzsprung-Russel diagram. Given the conclusions reached by \citet{rosu20a, rosu22a, rosu22b} about the necessity to include enhanced turbulent mixing in the stellar evolution models to reproduce the internal structure of the massive stars they studied, it is highly unlikely that a model without any turbulent diffusion would be representative of the stars of HD\,165052. Hence, we further discarded the tracks having $D_T=0$\,cm$^2$\,s$^{-1}$. We were thus left with nine tracks for the primary star and seven tracks for the secondary star.    

For each selected evolutionary track, we report in Table\,\ref{table:cles_models} the evolutionary mass, radius, effective temperature, and internal stellar structure constant corrected for the effects of rotation following \citet{claret99} at the ages of 1.5, 2.0, and 2.5\,Myr. The comparison between models of same initial mass and same current age shows us that the turbulent diffusion coefficient has a negligible impact on the current mass of the model and has only a small impact on the current radius and internal stellar structure constant of the model. This behaviour is not surprising given that the star is very young and, hence, the turbulent mixing occurring in its interior did not act for a long enough time to produce important changes to both the radius and the internal stellar structure constant. 

\begin{table*}
\caption{Properties of stellar structure and evolution models for the primary and secondary stars of HD\,165052 at the ages of 1.5, 2.0, and 2.5\,Myr. Column\,1 gives the name of the model. Columns\,2 and 3 give the initial mass $M_\text{init}$ and the turbulent diffusion coefficient $D_T$ of the evolutionary track. Columns\,4, 7, and 10 give the current mass $M$ of the model. Columns\,5, 8, and 11 give the current radius $R$ of the model. Columns\,6, 9, and 12 give the internal stellar structure constant $k_2$ of the model corrected for the effects of rotation according to \citet{claret99}.}
\label{table:cles_models}
\centering
\begin{tabular}{l c c | c c c | c c c | c c c}
\hline\hline
\vspace{-3mm}\\
\multicolumn{3}{c}{Evolutionary track} & \multicolumn{3}{c}{Age = 1.5\,Myr} & \multicolumn{3}{c}{Age = 2.0\,Myr} & \multicolumn{3}{c}{Age = 2.5\,Myr} \\
Name & $M_\text{init}$ & $D_T$ & $M$  & $R$ & $k_2$ & $M$ & $R$ & $k_2$ & $M$ & $R$ & $k_2$\\
 & ($M_\odot$) & (cm$^2$\,s$^{-1}$) & ($M_\odot$) & ($R_\odot$) & ($10^{-2}$) & ($M_\odot$) & ($R_\odot$) & ($10^{-2}$) & ($M_\odot$) & ($R_\odot$) & ($10^{-2}$)\\
\vspace*{-3mm}\\
\hline
\vspace*{-3mm}\\
\multicolumn{11}{l}{Primary star} \\
\vspace*{-3mm}\\
PM24DT1 & 24 & $1\times 10^7$ & 23.86 & 6.68 & 1.30 & 23.81 & 6.87 & 1.20 & 23.75 & 7.07 & 1.09 \\
PM24DT2 & 24 & $2\times 10^7$ & 23.86 & 6.64 & 1.32 & 23.81 & 6.81 & 1.22 & 23.75 & 6.97 & 1.12 \\
PM24DT3 & 24 & $3\times 10^7$ & 23.86 & 6.61 & 1.33 & 23.81 & 6.76 & 1.23 & 23.75 & 6.90 & 1.15 \\
PM25DT1 & 25 & $1\times 10^7$ & 24.83 & 6.87 & 1.29 & 24.77 & 7.07 & 1.18 & 24.69 & 7.29 & 1.06 \\
PM25DT2 & 25 & $2\times 10^7$ & 24.83 & 6.83 & 1.31 & 24.77 & 7.00 & 1.20 & 24.69 & 7.19 & 1.09 \\
PM25DT3 & 25 & $3\times 10^7$ & 24.83 & 6.79 & 1.32 & 24.76 & 6.95 & 1.22 & 24.69 & 7.12 & 1.12 \\
PM26DT1 & 26 & $1\times 10^7$ & 25.80 & 7.05 & 1.28 & 25.72 & 7.27 & 1.16 & 25.63 & 7.52 & 1.04 \\
PM26DT2 & 26 & $2\times 10^7$ & 25.80 & 7.01 & 1.29 & 25.72 & 7.21 & 1.18 & 25.62 & 7.41 & 1.06 \\
PM26DT3 & 26 & $3\times 10^7$ & 25.80 & 6.97 & 1.31 & 25.72 & 7.15 & 1.19 & 25.62 & 7.33 & 1.09 \\
\vspace*{-3mm}\\
\hline
\vspace*{-3mm}\\
\multicolumn{11}{l}{Secondary star} \\
\vspace*{-3mm}\\
SM20DT3 & 20 & $3\times 10^7$ & 19.95 & 5.86 & 1.35 & 19.93 & 5.96 & 1.29 & 19.90 & 6.05 & 1.23 \\ 
SM21DT1 & 21 & $1\times 10^7$ & 20.93 & 6.11 & 1.33 & 20.90 & 6.25 & 1.24 & 20.87 & 6.39 & 1.16 \\
SM21DT2 & 21 & $2\times 10^7$ & 20.93 & 6.08 & 1.34 & 20.90 & 6.20 & 1.26 & 20.87 & 6.32 & 1.19 \\
SM21DT3 & 21 & $3\times 10^7$ & 20.93 & 6.05 & 1.35 & 20.90 & 6.16 & 1.28 & 20.87 & 6.26 & 1.21 \\
SM22DT1 & 22 & $1\times 10^7$ & 21.91 & 6.31 & 1.32 & 21.88 & 6.46 & 1.23 & 21.84 & 6.62 & 1.14 \\
SM22DT2 & 22 & $2\times 10^7$ & 21.91 & 6.27 & 1.33 & 21.88 & 6.40 & 1.25 & 21.84 & 6.54 & 1.17 \\
SM22DT3 & 22 & $3\times 10^7$ & 21.91 & 6.24 & 1.34 & 21.88 & 6.36 & 1.27 & 21.84 & 6.48 & 1.19 \\
\vspace*{-3mm}\\
\hline
\end{tabular}
\end{table*}

Assuming that the primary model with $M_\text{init}=25\,M_\odot$ and $D_T=2\times 10^7$\,cm$^2$\,s$^{-1}$ and the secondary model with $M_\text{init}=21\,M_\odot$ and $D_T=2\times 10^7$\,cm$^2$\,s$^{-1}$, both at ages 2\,Myr, are representative of the stars, we infer evolutionary masses $M_\text{ev,P}=24.8\pm1.0\,M_\odot$ and $M_\text{ev,S}=20.9\pm1.0\,M_\odot$, where the subscript P and S stand for the primary and secondary stars, respectively, evolutionary radii $R_\text{ev,P}=7.0^{+0.5}_{-0.4}\,R_\odot$ and $R_\text{ev,S}=6.2^{+0.4}_{-0.3}\,R_\odot$, and internal stellar structure constants corrected for the stellar rotation following \citet{claret99} $k_{2,\text{P}}= 1.20^{+0.13}_{-0.16} \times 10^{-2}$ and $k_{2,\text{S}}= 1.26^{+0.09}_{-0.12} \times 10^{-2}$, where the angular rotation velocities $\Omega$ were computed using the projected rotational velocities derived in Sect.\,\ref{subsect:vsini} corrected for the inclination.  The error bars include the differences coming from a difference of $1.0\,M_\odot$ in $M_\text{init}$, of $1\times10^7$\,cm$^2$\,s$^{-1}$ in $D_T$, and of 0.5\,Myr in the stellar age. The evolutionary mass ratio of the binary amounts to $0.84^{+0.08}_{-0.07}$, slightly smaller than but still within the error bars of the observational mass ratio. Combined with the minimum masses obtained in Sect.\,\ref{sect:omegadot} (see last column of Table\,\ref{table:RVs_fit}), the evolutionary masses put a constraint on the orbital inclination: $22.1^\circ \le i \le 23.3^\circ$. This small value of the inclination indicates that eclipses are very unlikely to be seen in photometric observations of this binary system (see also Fig.\,\ref{fig:lightcurve}).

We then computed theoretical values for the apsidal motion rate. We adopted models for the primary and secondary stars of the same age and combined all the possible pairs given in Table\,\ref{table:cles_models}. The semi-major axis is computed through the Kepler's third law for the corresponding combination of primary and secondary masses whilst the rotational periods of the stars are computed using the projected rotational velocities derived in Sect.\,\ref{subsect:vsini} corrected for the inclination. The results are provided in Table\,\ref{table:cles_omegadot}: We adopted the models having an age of 2.0\,Myr as our reference models and computed the error bars based on models of 1.5 and 2.5\,Myr. All theoretical apsidal motion rates agree with the observational determination (see last column in Table\,\ref{table:RVs_fit}) except those computed combining the pairs of models PM24DT2-SM20DT3 and PM24DT3-SM20DT3 that slightly underestimate the apsidal motion rate compared to the observational value. This confrontation between observational and theoretical apsidal motion rates allows us to confirm the inferred evolutionary masses and radii of the stars. 

\begin{table*}
\caption{Theoretical values of the apsidal motion rate (in $^\circ$\,yr$^{-1}$) in HD\,165052. The values are obtained with models for the primary and secondary stars of 2.0\,Myr and the error bars are computed using models of 1.5 and 2.5\,Myr.}
\label{table:cles_omegadot}
\centering
\begin{tabular}{l | l l l l l l l}
\hline\hline
& SM20DT3 & SM21DT1 & SM21DT2 & SM21DT3 & SM22DT1 & SM22DT2 & SM22DT3 \\ 
\hline
\vspace*{-3mm}\\
PM24DT1 & $10.54^{+0.34}_{-0.36}$ & $11.13^{+0.43}_{-0.45}$ & $11.00^{+0.38}_{-0.40}$ & $10.92^{+0.35}_{-0.38}$ & $11.52^{+0.47}_{-0.46}$ & $11.38^{+0.41}_{-0.41}$ & $11.29^{+0.37}_{-0.39}$ \\
\vspace*{-3mm}\\
PM24DT2 & $10.41^{+0.28}_{-0.31}$ & $10.99^{+0.37}_{-0.40}$ & $10.87^{+0.32}_{-0.36}$ & $10.78^{+0.29}_{-0.33}$ & $11.39^{+0.41}_{-0.41}$ & $11.25^{+0.34}_{-0.37}$ & $11.15^{+0.31}_{-0.34}$ \\
\vspace*{-3mm}\\
PM24DT3 & $10.32^{+0.25}_{-0.28}$ & $10.90^{+0.34}_{-0.37}$ & $10.78^{+0.28}_{-0.33}$ & $10.69^{+0.26}_{-0.30}$ & $11.30^{+0.37}_{-0.38}$ & $11.15^{+0.31}_{-0.34}$ & $11.06^{+0.28}_{-0.31}$ \\
\vspace*{-3mm}\\
PM25DT1 & $10.85^{+0.38}_{-0.38}$ & $11.44^{+0.46}_{-0.47}$ & $11.31^{+0.41}_{-0.42}$ & $11.22^{+0.38}_{-0.40}$ & $11.83^{+0.50}_{-0.48}$ & $11.69^{+0.44}_{-0.43}$ & $11.59^{+0.41}_{-0.41}$ \\
\vspace*{-3mm}\\
PM25DT2 & $10.71^{+0.31}_{-0.32}$ & $11.29^{+0.39}_{-0.41}$ & $11.16^{+0.34}_{-0.37}$ & $11.08^{+0.31}_{-0.35}$ & $11.68^{+0.43}_{-0.43}$ & $11.54^{+0.37}_{-0.38}$ & $11.45^{+0.34}_{-0.35}$ \\
\vspace*{-3mm}\\
PM25DT3 & $10.61^{+0.26}_{-0.29}$ & $11.20^{+0.35}_{-0.38}$ & $11.07^{+0.30}_{-0.34}$ & $10.98^{+0.27}_{-0.32}$ & $11.59^{+0.38}_{-0.40}$ & $11.45^{+0.32}_{-0.35}$ & $11.35^{+0.29}_{-0.32}$ \\
\vspace*{-3mm}\\
PM26DT1 & $11.18^{+0.40}_{-0.40}$ & $11.76^{+0.49}_{-0.49}$ & $11.63^{+0.44}_{-0.45}$ & $11.54^{+0.41}_{-0.42}$ & $12.14^{+0.52}_{-0.50}$ & $12.00^{+0.46}_{-0.46}$ & $11.91^{+0.43}_{-0.43}$ \\
\vspace*{-3mm}\\
PM26DT2 & $11.02^{+0.33}_{-0.35}$ & $11.60^{+0.41}_{-0.44}$ & $11.47^{+0.36}_{-0.39}$ & $11.39^{+0.33}_{-0.37}$ & $11.99^{+0.45}_{-0.45}$ & $11.84^{+0.39}_{-0.40}$ & $11.75^{+0.35}_{-0.37}$ \\
\vspace*{-3mm}\\
PM26DT3 & $10.91^{+0.28}_{-0.31}$ & $11.49^{+0.37}_{-0.40}$ & $11.36^{+0.32}_{-0.35}$ & $11.28^{+0.29}_{-0.33}$ & $11.88^{+0.40}_{-0.41}$ & $11.74^{+0.34}_{-0.36}$ & $11.64^{+0.31}_{-0.34}$\\
\hline
\end{tabular}
\end{table*}

Finally, we computed the observational weighted-average mean of the internal structure constants of the stars, $k_{2,\text{obs}}$, as defined by Equations\,(18) and (19) in \citet{rosu22a}, adopting an inclination of $22.7^\circ \pm 0.6^\circ$ while all other parameters are taken from observational determinations. We obtained $k_{2,\text{obs}} = 1.36^{+0.44}_{-0.43} \times 10^{-2}$, a value slightly larger but still compatible with the theoretical determinations for the two stars within the error bars.

\section{Conclusion}
\label{sect:conclusion}
We presented a new, in depth spectroscopic analysis of medium- and high-resolution spectra of the massive eccentric binary system HD\,165052 and derived the first self-consistent orbital solution of all existing RV data, including those reported in the literature, accounting for the change of the longitude of periastron with time.

We applied our disentangling code based on the method of \citet{GL} to derive the RVs of the stars at each time of observations. Then, we applied the advanced disentangling method proposed by \citet{quintero20}, that allows the reconstruction of artefact-free individual spectra, to the spectroscopic observations to reconstruct the spectra of the components. These latter were analysed, for the first time, with the non-LTE model atmosphere code \texttt{CMFGEN} to derive fundamental stellar and wind parameters. 

We performed the RV analysis of all data, including those coming from the literature, along three avenues: considering the primary RVs only, the secondary RVs only, and converting the primary and secondary RVs into primary-equivalent RVs. In all three cases, we obtained orbital solutions which were in agreement within their error bars. We adopted the equivalent RVs solution as the final solution and hence, concluded that the apsidal motion rate in the system amounts to $(11.30^{+0.64}_{-0.49})^\circ$\,yr$^{-1}$. We conclude that our analysis of the spectroscopic observations of HD\,165052 and the  analysis of the RVs of the stars explicitly accounting for the apsidal motion allow us to derive a consistent measure of the apsidal motion rate in the binary. Our analysis of all available RV data also allowed us to determine more reliably the orbital parameters of the binary. Our values differ from those of \citet{stickland97}, \citet{arias02}, and \citet{linder07} mainly because we explicitly accounted for the apsidal motion in the RV analysis. Our results differ from those of \citet{FER} mainly due to the increased number of RV data available at the time of our study.

We computed dedicated stellar structure and evolution tracks with \texttt{Cl\'es} assuming the stellar effective temperatures and luminosities obtained as part of the \texttt{CMFGEN} analysis, and an age of $2.0\pm 0.5$\,Myr for the binary. We derived evolutionary masses $M_\text{ev,P}=24.8\pm1.0\,M_\odot$ and $M_\text{ev,S}=20.9\pm1.0\,M_\odot$ and evolutionary radii $R_\text{ev,P}=7.0^{+0.5}_{-0.4}\,R_\odot$ and $R_\text{ev,S}=6.2^{+0.4}_{-0.3}\,R_\odot$ for the primary and secondary stars, respectively. Through the confrontation with the minimum stellar masses, we constrained the orbital inclination: $22.1^\circ \le i \le 23.3^\circ$. We computed theoretical apsidal motion rates adopting  two stellar models of the same age and observed that these were in agreement with the observational apsidal motion rate, therefore enforcing the inferred values of the evolutionary stellar masses and radii, and putting some constraints on the density stratification inside the stars through the internal stellar structure constants.

\begin{figure}
\includegraphics[clip=true, trim=10 40 20 270, width=\linewidth]{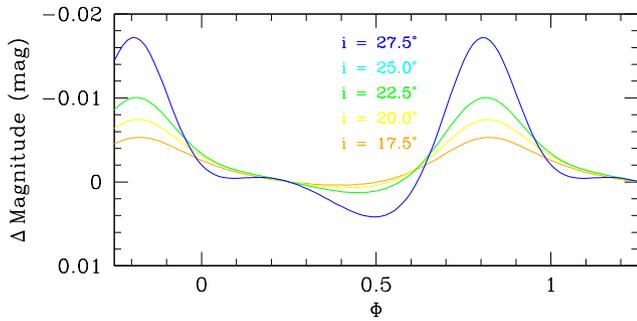}
\caption{Synthetic lightcurves of HD\,165052 for different values of the orbital inclination between $17.5^{\circ}$ and $27.5^{\circ}$. We note that the exact shape of the lightcurves depends on the $\omega$-value, hence on the assumed observational date; in the present case we fixed the date to HJD 2\,460\,000.}
\label{fig:lightcurve}
\end{figure}

Whilst our results exclude the possibility of photometric eclipses, it would nevertheless be worth to acquire high-precision space-borne photometry to look for low-level photometric variability due to a phase-dependent tidal distortion of the stars and/or tidally-induced pulsations \citep[e.g.,][]{Kolaczek21}. Adopting our best estimates of the stellar parameters, we used the Nightfall code to simulate synthetic lightcurves of HD\,165052 for different values of the orbital inclination between $17.5^{\circ}$ and $27.5^{\circ}$ (see Fig.\,\ref{fig:lightcurve}).  In all cases, we found that the lightcurve is dominated by the so-called heartbeat variations due to the orbital eccentricity rather than by ellipsoidal variations. For the most likely inclination near $22.5^{\circ}$, we expect peak-to-peak amplitudes of about 0.01\,mag which are certainly within reach of sensitive space-borne photometry. We inspected both ASAS and Hipparcos photometric data. No significative variation is observed for the ASAS lightcurve and the error bars on the data are larger than the expected variations. Whilst the Hipparcos data have lower error bars than the ASAS data, the errors remain quite large and the data do not sample the most interesting parts of the orbit, namely the phases close to periastron passage. Unfortunately, there exist currently no other high-precision space-borne photometric data. Hence, the existing data cannot be used to study the light curve of this system. Whereas heartbeat variations could help confirming the estimate of the orbital inclination that we have obtained, the detection of pulsations would offer the possibility to gain further insight into the internal structure of the stars that would offer a powerful tool for comparison with the \texttt{Cl\'es} models that we have tested here. 

\section*{Acknowledgements}
S. Rosu acknowledges support from the Fonds de la Recherche Scientifique (F.R.S.- FNRS, Belgium). We thank Dr John Hillier for making his code {\tt CMFGEN} publicly available. This research uses optical spectra collected with the TIGRE telescope (La Luz, Mexico). TIGRE is a collaboration of Hamburger Sternwarte, the Universities of Hamburg, Guanajuato, and Li\`ege. The authors thank the referee for his/her suggestions and comments towards the improvement of the manuscript.

\section*{Data Availability}
The ESO and CFHT data used in this study are available from the corresponding archives, whilst the TIGRE spectra can be made available upon reasonable request.

\appendix
\section{Spectral disentangling based on the method described by \citet{GL}}
\label{appendix:spectrotable}
This appendix provides the journal of the spectroscopic observations of HD\,165052 (Table\,\ref{Table:spectro+RV}) and a detailed description of the  spectral disentangling performed to derive the RVs of the stars.

For the disentangling, we used synthetic {\tt TLUSTY} spectra with $T_\text{eff} = 35\,000$\,K, $\log g = 4.0 $, and $v \sin i_\text{rot} =  70$\,km\,s$^{-1}$ as cross-correlation templates in the determination of the RVs.

We performed the disentangling on 15 separate wavelength domains: B1[3810:4150]\,\AA, B2[4150:4250]\,\AA, B3[4300:4570]\,\AA, B4[4600:5040]\,\AA, B5[4600:4980]\,\AA, B6[4800:5040]\,\AA, B7[5380:5610]\,\AA, B8[5380:5840]\,\AA, B9[5790:5840]\,\AA, B10[5380:5450]\,\AA, B11[5380:5750]\,\AA, and B12[5560:5840]\,\AA~in the blue domain, and R1[5860:5885]\,\AA, R2[6500:6700]\,\AA, and R3[7000:7100]\,\AA~in the red domain. As a first step, we only considered the FEROS, ESPaDOns, and TIGRE spectra as these spectra have a better resolution than the UVES and XSHOOTER spectra. We processed the wavelength domains (B1, B2, B3, B4, B7, B8, B9, R1, R2, and R3) to reproduce the individual spectra and simultaneously estimate the RVs of the stars. The TIGRE spectra cover all above-mentioned domains except for B8, and three FEROS spectra (taken at 2\,451\,304.7434, 2\,451\,304.7507, and 2\,451\,304.9309 HJD) do not cover the B2 domain due to instrumental artefacts in that wavelength domain. We averaged the RVs from the individual wavelength domains using weights corresponding to the number of strong lines present in these domains (five lines for B1, one for B2, three for B3, two for B4, two for B7, three for B8, two for B9, one for R1, one for R2, and one for R3). The resulting RVs of both stars are reported in Table\,\ref{Table:spectro+RV} together with their $1\sigma$ errors. We then performed the disentangling on the ten domains covered by the XSHOOTER and UVES spectra (B1, B2, B3, B5, B6, B10, B11, B12, R2, and R3) with the RVs of XSHOOTER and UVES observations free to vary, and the RVs of the FEROS, ESPaDOns, and TIGRE spectra fixed to their previously computed weighted averages. Two UVES observations (taken at 2\,452\,584.5002 and 2\,452\,584.5161 HJD) cover the B1, B2, B3, and B5 domains only, the other two UVES observations (taken at 2\,452\,584.5216 and 2\,452\,584.5234 HJD) cover the B6, B10, B11, and R2 domains only, four XSHOOTER observations (taken at 2\,457\,583.7901, 2\,457\,583.7916, 2\,457\,954.7756, and 2\,457\,954.7772 HJD) cover the B1, B2, B3, B5, B6, and B10 domains only, and the remaining five XSHOOTER observations (taken at 2\,457\,583.7902, 2\,457\,583.7916, 2\,457\,583.8401, 2\,457\,954.7757, and 2\,457\,954.7773 HJD) cover the B12, R2, and R3 domains only. We averaged the RVs from the individual wavelength domains using weights corresponding to the number of strong lines present in these domains (five lines for B1, one for B2, three for B3, two for B5, one for B6, one for B10, two for B11, two for B12, one for R2, and one for R3). The resulting RVs of both stars are reported in Table\,\ref{Table:spectro+RV} together with their $1\sigma$ errors.

\begin{table*}
\caption{Journal of the spectroscopic observations of HD\,165052.}
\label{Table:spectro+RV}
\centering
\begin{tabular}{l l l l l}
\hline
\hline
\vspace{-3mm}\\
HJD & $\phi$  & $\text{RV}_\text{P}$ & $\text{RV}_\text{S}$ & Instrumentation\\
--\,2\,450\,000&  & (km\,s$^{-1}$) & (km\,s$^{-1}$) & \\
\hline
\vspace{-3mm}\\
1299.7250 &0.236& $92.8\pm1.1$ & $-101.8\pm1.2$ & ESO 1.5\,m + FEROS\\
1300.7319 &0.577& $-61.3\pm1.5$ & $76.4\pm0.7$ & ESO 1.5\,m + FEROS\\
1300.9264 &0.643& $-81.5\pm1.2$ & $97.5\pm0.7$ & ESO 1.5\,m + FEROS\\
1301.9281 &0.982& $30.5\pm0.6$ & $-26.0\pm1.7$ & ESO 1.5\,m + FEROS\\
1304.7434 &0.934& $4.5\pm0.7$ & $0.2\pm0.9$ & ESO 1.5\,m + FEROS\\
1304.7507 &0.937& $11.5\pm0.3$ & $-6.5\pm1.2$ & ESO 1.5\,m + FEROS\\
1304.9309 &0.998& $40.3\pm0.4$ & $-38.4\pm1.4$ & ESO 1.5\,m + FEROS\\
1323.8361 &0.393& $24.9\pm1.3$ & $-23.8\pm2.2$ & ESO 1.5\,m + FEROS\\
1327.6014 &0.667& $-88.6\pm1.1$ & $101.5\pm0.6$ & ESO 1.5\,m + FEROS\\
1327.9127 &0.773& $-88.8\pm1.0$ & $99.4\pm0.6$ & ESO 1.5\,m + FEROS\\
1670.7102 &0.745& $-83.0\pm0.9$ & $99.2\pm0.7$ & ESO 1.5\,m + FEROS\\
1671.7197 &0.087& $103.3\pm1.1$ & $-108.4\pm1.7$ & ESO 1.5\,m + FEROS\\
1672.7016 &0.419& $-9.5\pm1.2$ & $16.4\pm0.9$ & ESO 1.5\,m + FEROS\\
2335.8879 &0.783& $-46.2\pm1.5$ & $57.7\pm0.8$ & ESO 1.5\,m + FEROS\\
2336.8791 &0.118& $105.5\pm0.7$ & $-113.9\pm1.1$ & ESO 1.5\,m + FEROS\\
2337.8880 &0.460& $-54.3\pm1.0$ & $66.8\pm0.6$ & ESO 1.5\,m + FEROS\\
2338.8808 &0.795& $-39.3\pm1.2$ & $48.8\pm0.6$ & ESO 1.5\,m + FEROS\\
2339.8848 &0.135& $101.1\pm0.6$ & $-109.2\pm0.9$ & ESO 1.5\,m + FEROS\\
2381.7324 &0.293& $ 5.6\pm1.0$ & $-2.4\pm1.3$ & ESO 1.5\,m + FEROS\\
2382.8570 &0.673& $-81.7\pm0.8$ & $96.7\pm0.6$ & ESO 1.5\,m + FEROS\\
2383.8577 &0.012& $95.5\pm0.7$ & $-101.6\pm1.1$ & ESO 1.5\,m + FEROS\\
2584.5002 &0.891& $54.1\pm1.6$ & $-10.0\pm1.2$ & ESO VLT + UVES\\
2584.5161 &0.897& $49.7\pm1.4$ & $-7.0\pm0.5$ & ESO VLT + UVES\\
2584.5216 &0.899& $55.8\pm2.5$ & $-10.1\pm1.2$ & ESO VLT + UVES\\
2584.5234 &0.899& $55.7\pm2.6$ & $-11.4\pm1.7$ & ESO VLT + UVES\\
3130.9089 &0.748& $-25.2\pm1.2$ & $31.9\pm0.7$ & ESO 2.2\,m + FEROS\\
3134.9113 &0.102& $93.6\pm0.8$ & $-101.0\pm1.0$ & ESO 2.2\,m + FEROS\\
4212.8381 &0.778& $48.3\pm0.3$ & $-48.7\pm1.0$ & ESO 2.2\,m + FEROS\\
5351.9169 &0.142& $-39.3\pm1.2$ & $49.3\pm0.8$ & CFHT 3.6\,m + ESPaDOns\\
5351.9169 &0.142& $-39.3\pm1.2$ & $49.4\pm0.8$ & CFHT 3.6\,m + ESPaDOns\\
5351.9302 &0.147& $-41.6\pm1.2$ & $52.2\pm0.8$ & CFHT 3.6\,m + ESPaDOns\\
5351.9302 &0.147& $-41.7\pm1.3$ & $52.2\pm0.8$ & CFHT 3.6\,m + ESPaDOns\\
5351.9368 &0.149& $-43.4\pm1.2$ & $53.1\pm0.8$ & CFHT 3.6\,m + ESPaDOns\\
5351.9368 &0.149& $-43.4\pm1.2$ & $53.1\pm0.8$ & CFHT 3.6\,m + ESPaDOns\\
5351.9434 &0.151& $-44.4\pm1.2$ & $55.1\pm0.8$ & CFHT 3.6\,m + ESPaDOns\\
5351.9434 &0.151& $-44.3\pm1.1$ & $55.1\pm0.8$ & CFHT 3.6\,m + ESPaDOns\\
5351.9567 &0.156& $-47.0\pm1.1$ & $57.9\pm0.9$ & CFHT 3.6\,m + ESPaDOns\\
5351.9568 &0.156& $-46.9\pm1.1$ & $57.9\pm0.9$ & CFHT 3.6\,m + ESPaDOns\\
7583.7901 &0.212& $-83.2\pm0.6$ & $107.8\pm0.7$ & ESO VLT + XSHOOTER\\
7583.7902 &0.212& $-77.3\pm1.8$ & $114.4\pm1.9$ & ESO VLT + XSHOOTER\\
7583.7916 &0.213& $-83.2\pm0.7$ & $107.2\pm0.7$ & ESO VLT + XSHOOTER\\
7583.7916 &0.213& $-74.3\pm2.4$ & $112.8\pm1.5$ & ESO VLT + XSHOOTER\\
7583.8401 &0.229& $-73.8\pm1.9$ & $105.0\pm2.5$ & ESO VLT + XSHOOTER\\
7954.7756 &0.721& $91.1\pm0.7$ & $-78.0\pm0.8$ & ESO VLT + XSHOOTER\\
7954.7757 &0.721& $94.1\pm2.5$ & $-72.7\pm2.2$ & ESO VLT + XSHOOTER\\
7954.7772 &0.721& $90.7\pm0.7$ & $-78.2\pm0.9$ & ESO VLT + XSHOOTER\\
7954.7773 &0.721& $94.4\pm2.9$ & $-72.2\pm2.3$ & ESO VLT + XSHOOTER\\
8574.9673 &0.539& $87.4\pm1.2$ & $-99.1\pm0.8$ & TIGRE + HEROS\\
8577.9071 &0.534& $88.8\pm1.2$ & $-100.0\pm1.0$ & TIGRE + HEROS\\
8577.9458 &0.547& $85.3\pm1.4$ & $-98.1\pm1.1$ & TIGRE + HEROS\\
8580.8998 &0.546& $87.9\pm1.6$ & $-101.5\pm0.8$ & TIGRE + HEROS\\
8583.8868 &0.557& $90.7\pm1.9$ & $-99.4\pm0.9$ & TIGRE + HEROS\\
8595.8649 &0.609& $82.2\pm1.5$ & $-94.2\pm1.4$ & TIGRE + HEROS\\
8620.8763 &0.071& $-104.3\pm1.0$ & $114.7\pm2.1$ & TIGRE + HEROS\\
8622.7862 &0.717& $50.8\pm0.8$ & $-59.0\pm1.4$ & TIGRE + HEROS\\
\hline
\end{tabular}
\end{table*}
\setcounter{table}{0}

\begin{table*}
\caption{continued.}
\centering
\begin{tabular}{l l l l l}
\hline
\hline
\vspace{-3mm}\\
HJD & $\phi$  & $\text{RV}_\text{P}$ & $\text{RV}_\text{S}$ & Instrumentation\\
--\,2\,450\,000&  & (km\,s$^{-1}$) & (km\,s$^{-1}$) & \\
\hline
\vspace{-3mm}\\
8623.7795 &0.053& $-103.4\pm1.4$ & $118.0\pm0.8$ & TIGRE + HEROS\\
8625.7823 &0.731& $47.1\pm1.3$ & $-51.1\pm1.2$ & TIGRE + HEROS\\
8626.8430 &0.089& $-99.6\pm1.1$ & $110.1\pm3.0$ & TIGRE + HEROS\\
8943.9344 &0.365& $55.3\pm1.0$ & $-64.5\pm1.4$ & TIGRE + HEROS\\
8953.8765 &0.729& $29.4\pm1.5$ & $-34.5\pm1.1$ & TIGRE + HEROS\\
8962.9326 &0.793& $3.2\pm1.1$ & $-1.9\pm1.0$ & TIGRE + HEROS\\
8971.9508 &0.844& $-37.4\pm1.5$ & $39.2\pm0.9$ & TIGRE + HEROS\\
8981.8337 &0.187& $-42.2\pm1.3$ & $43.9\pm1.0$ & TIGRE + HEROS\\
8990.9362 &0.267& $-15.5\pm1.4$ & $13.0\pm1.3$ & TIGRE + HEROS\\
8994.8576 &0.593& $79.0\pm1.8$ & $-91.2\pm2.1$ & TIGRE + HEROS\\
9000.8744 &0.629& $71.0\pm1.5$ & $-79.8\pm1.5$ & TIGRE + HEROS\\
9014.8052 &0.342& $48.4\pm1.8$ & $-56.5\pm1.4$ & TIGRE + HEROS\\
9035.7900 &0.441& $79.6\pm2.1$ & $-92.2\pm1.7$ & TIGRE + HEROS\\
9045.7161 &0.799& $3.9\pm1.2$ & $-3.4\pm0.8$ & TIGRE + HEROS\\
9297.9856 &0.145& $-49.8\pm1.2$ & $55.8\pm0.9$ & TIGRE + HEROS\\
9311.9249 &0.861& $-64.5\pm1.3$ & $70.1\pm1.1$ & TIGRE + HEROS\\
\hline
\end{tabular}
\begin{tablenotes}
\item Column\,1 gives the heliocentric Julian date (HJD) of the observations at mid-exposure. Column\,2 gives the observational phase $\phi$ computed with the orbital period determined in Sect.\,\ref{sect:omegadot} based on the equivalent RVs (last column in Table\,\ref{table:RVs_fit}). Columns\,3 and 4 give the radial velocities $\text{RV}_\text{P}$ and $\text{RV}_\text{S}$ of the primary and secondary stars, respectively. The errors represent $\pm1\sigma$. Column\,5 provides information about the instrumentation.
\end{tablenotes}
\end{table*}

\section{Journal of the RVs of HD\,165052 coming from \citet{FER}}
\label{appendix:RVs_ferrero}
This appendix provides the journal of the RVs of HD\,165052 from \citet{FER}. Given that the authors do not provide any error bar on their mean RVs, we computed both the mean RVs and standard deviation based on their individual measurements given in their Table B2, which we provide in Table\,\ref{table:RVs_ferrero}.

\begin{table}
\caption{Journal of the RVs of HD\,165052 from \citet{FER}.}
\label{table:RVs_ferrero}
\centering
\begin{tabular}{l l l}
\hline
\hline
\vspace{-3mm}\\
HJD  & $\text{RV}_\text{P}$ & $\text{RV}_\text{S}$ \\
--\,2\,450\,000&  (km\,s$^{-1}$) & (km\,s$^{-1}$) \\
\hline
\vspace{-3mm}\\
4582.8675 & $98.3\pm8.4$ & $-106.4\pm6.4$ \\
4693.6231 & $-80.9\pm10.3$ & $90.8\pm11.8$ \\
4695.6483 & $15.6\pm10.5$ & $-13.2\pm8.6$ \\
4696.5568 & $-83.3\pm8.8$ & $94.2\pm5.6$ \\
4696.6286 & $-79.4\pm7.2$ & $93.9\pm8.1$ \\
4696.6605 & $-77.3\pm8.5$ & $91.6\pm6.7$ \\
4697.6220 & $62.9\pm6.3$ & $-76.0\pm5.9$ \\
4955.8004 & $-7.0\pm14.6$ & $9.6\pm10.6$ \\
4956.9100 & $-57.0\pm8.6$ & $62.4\pm8.4$ \\
4964.8865 & $-56.1\pm11.4$ & $69.9\pm8.4$ \\
4965.8323 & $-47.3\pm12.0$ & $56.7\pm11.1$ \\
4966.8550 & $100.8\pm6.2$ & $-106.9\pm8.4$ \\
4967.8220 & $-55.7\pm14.9$ & $65.5\pm12.4$ \\
4968.8723 & $-38.3\pm8.8$ & $41.3\pm6.0$ \\
5046.6533 & $102.5\pm7.1$ & $-109.5\pm8.2$ \\
5047.7305 & $-67.7\pm13.6$ & $85.2\pm9.4$ \\
5048.7185 & $-26.2\pm9.7$ & $30.2\pm6.8$ \\
5049.7159 & $101.7\pm10.5$ & $-107.7\pm5.6$ \\
5052.7320 & $100.1\pm6.4$ & $-103.5\pm5.7$ \\
5337.6475 & $-94.2\pm9.3$ & $101.4\pm8.5$ \\
5339.7307 & $44.5\pm5.5$ & $-50.1\pm6.1$ \\
5340.6327 & $-95.7\pm8.3$ & $101.9\pm11.8$ \\
5341.5812 & $30.4\pm14.1$ & $-33.0\pm10.6$ \\
5342.6292 & $54.6\pm11.7$ & $-63.8\pm10.2$ \\
5376.5522 & $-52.3\pm12.1$ & $59.7\pm9.9$ \\
5378.7551 & $-77.5\pm11.1$ & $89.5\pm9.2$ \\
5380.5253 & $94.3\pm9.2$ & $-104.5\pm8.0$ \\
5381.7473 & $-82.9\pm9.7$ & $91.3\pm7.0$ \\
5383.7515 & $90.5\pm6.0$ & $-97.7\pm6.3$ \\
5429.6974 & $-58.2\pm11.8$ & $68.6\pm7.4$ \\
5430.6168 & $87.2\pm7.9$ & $-91.1\pm8.0$ \\
5431.6959 & $-42.3\pm11.7$ & $51.6\pm7.5$ \\
5432.6362 & $-55.9\pm10.9$ & $69.1\pm7.4$ \\
5433.5059 & $78.7\pm11.4$ & $-88.0\pm7.3$ \\
5434.6834 & $-45.7\pm14.0$ & $60.3\pm7.5$ \\
5435.5627 & $-64.5\pm11.8$ & $71.7\pm9.3$ \\
5698.8397 & $-7.9\pm9.3$ & $41.7\pm8.3$ \\
\hline
\end{tabular}
\begin{tablenotes}
\item Column\,1 gives the heliocentric Julian date (HJD) of the observations at mid-exposure. Columns\,2 and 3 give the radial velocities $\text{RV}_\text{P}$ and $\text{RV}_\text{S}$ of the primary and secondary stars, respectively. The errors represent $\pm1\sigma$.
\end{tablenotes}
\end{table}

\bsp	
\label{lastpage}
\end{document}